\documentclass[12pt,a4]{article}

\usepackage{graphicx,amsmath,amsthm,bm,fancybox,ascmac,amssymb,multirow,hhline,atbegshi}

\usepackage{natbib}
\usepackage{verbatim}
\usepackage{color}

\makeatletter
\def\section{\@startsection {section}{1}{\z@}{-3.5ex plus -1ex minus -.2ex}{2.3 ex plus .2ex}{\large\sc\centering}}
\def\subsection{\@startsection {subsection}{1}{\z@}{-3.5ex plus -1ex minus -.2ex}{2.3 ex plus .2ex}{\large}}
\makeatother

\theoremstyle{definition}

\newtheorem{thm}{Theorem}
\newtheorem{lem}{Lemma}

\newtheorem{coro}{Corollary}
\newcommand{\argmin}{\mathop{\rm argmin}}

\title{\large\bf AIC for the Non-concave Penalized Likelihood Method}
\author{\normalsize Yuta Umezu\thanks{Graduate School of Mathematics, Kyushu University. 744 Motooka, Nishi-ku, Fukuoka 819-0395, Japan}
\and  \normalsize Yusuke Shimizu$^*$
\and  \normalsize Hiroki Masuda$^*$
\and \normalsize Yoshiyuki Ninomiya\thanks{Corresponding author.
Institute of Mathematics for Industry, Kyushu University. 744 Motooka, Nishi-ku, Fukuoka 819-0395, Japan.
Email: nino@imi.kyushu-u.ac.jp}
}
\date{\normalsize Version: \today}

\selectfont

\allowdisplaybreaks

\begin{document}
\maketitle
\begin{abstract}
Non-concave penalized maximum likelihood methods, such as the Bridge, the SCAD, and the MCP, are widely used because they not only perform the parameter estimation and variable selection simultaneously but also are more efficient than the Lasso. They include a tuning parameter which controls a penalty level, and several information criteria have been developed for selecting it. While these criteria assure the model selection consistency, they have a problem in that there are no appropriate rules for choosing one from the class of information criteria satisfying such a preferred asymptotic property. In this paper, we derive an information criterion based on the original definition of the AIC by considering minimization of the prediction error rather than model selection consistency. Concretely speaking, we derive a function of the score statistic that is asymptotically equivalent to the non-concave penalized maximum likelihood estimator and then provide an estimator of the Kullback-Leibler divergence between the true distribution and the estimated distribution based on the function, whose bias converges in mean to zero. Furthermore, through simulation studies, we find that the performance of the proposed information criterion is about the same as or even better than that of the cross-validation.

\

\end{abstract}

\newpage

\section{Introduction}
The Lasso (\citealt{Tib96}) is a regularization method that imposes an $\ell_{1}$ penalty term $\lambda\|\bm{\beta}\|_{1}$ on an estimating function with respect to an unknown parameter vector $\bm{\beta}=(\beta_{1},\beta_{2},\ldots,\beta_{p})^{{\rm T}}$, where $\lambda\;(>0)$ is a tuning parameter controlling a penalty level. The Lasso can simultaneously perform estimation and variable selection by exploiting the non-differentiability of the penalty term at the origin. Concretely speaking, if $\hat{\bm{\beta}}_{\lambda}=(\hat{\beta}_{\lambda,1},\hat{\beta}_{\lambda,2},\ldots,\hat{\beta}_{\lambda,p})^{{\rm T}}$ is the estimator based on the Lasso, several of its components will shrink to exactly 0 when $\lambda$ is not close to 0. However, a parameter estimation based on the Lasso is not necessarily efficient, because the Lasso shrinks the estimator to the zero vector too much. To avoid such a problem, it has been proposed to use a penalty term that does not shrink the estimator with a large value. Typical examples of such regularization methods are the Bridge (\citealt{FraFri93}), the smoothly clipped absolute deviation (SCAD; \citealt{FanLi01}), and the minimax concave penalty (MCP; \citealt{Zha10}). Whereas the Bridge uses an $\ell_{q}$ penalty term ($0<q<1$), SCAD and MCP use penalty terms that can be approximated by an $\ell_{1}$ penalty term in the neighborhood of the origin, which we call an $\ell_{1}$ type. Although it is difficult to obtain estimates of them as their penalties are non-convex, there are several algorithms, such as the coordinate descent method and the gradient descent method that assure convergence to a local optimal solution.

On the other hand, in the above regularization methods, we have to choose a proper value for the tuning parameter $\lambda$, and this is an important task for appropriate model selection. One of the simplest ways of selecting $\lambda$ is to use cross-validation (CV; \citealt{Sto74}). While the stability selection method (\citealt{MeiBuh10}) based on subsampling in order to avoid problems caused by selecting a model based on only one value of $\lambda$ would be nice, it carries with it a considerable computational cost as in CV. Recently, information criteria without such a problem have been developed (\citealt{YuaLin07,WanLiTsa07,WanLiLen09,ZhaLiTsa10,FanTan13}). Here, by letting $\ell(\cdot)$ be the log-likelihood function and $\hat{\bm{\beta}}_{\lambda}$ be the estimator of $\bm{\beta}$ obtained by the above regularization methods, their information criteria take the form $-2\ell(\hat{\bm{\beta}}_{\lambda})+\kappa_{n}\|\hat{\bm{\beta}}_{\lambda}\|_{0}$. Accordingly, model selection consistency is at least assured for some sequence $\kappa_{n}$ that depends on at least the sample size $n$. For example, the information criterion with $\kappa_{n}=\log n$ is proposed as the BIC. This approach includes the results for the case in which the dimension of the parameter vector $p$ goes to infinity, and hence, it is considered to be significant. However, the choice of tuning parameter remains somewhat arbitrary. That is, there is a class of $\kappa_{n}$ assuring a preferred asymptotic property such as model selection consistency, but there are no appropriate rules for choosing one from the class. For example, since the BIC described above is not derived from the Bayes factor, there is no reason to use $\kappa_n=\log n$ instead of $\kappa_n=2\log n$. This is a severe problem because data analysts can choose $\kappa_{n}$ arbitrarily and do model selection as they want.

Information criteria without such an arbitrariness problem have been proposed by \citet{EfrHasJohTib04} or \citet{ZouHasTib07} for Gaussian linear regression and by \citet{NinKaw14} for generalized linear regression. Concretely speaking, on the basis of the original definition of the $C_{p}$ or AIC, they derive an unbiased estimator of the mean squared error or an asymptotically unbiased estimator of a Kullback-Leibler divergence. However, these criteria are basically only for the Lasso. In addition, the asymptotic setting used in \citet{NinKaw14} does not assure even estimation consistency.

Our goal in this paper is to derive an information criterion based on the original definition of AIC in an asymptotic setting that assures estimation consistency for regularization methods using non-concave penalties including the Bridge, SCAD, and MCP. To achieve it, the results presented in \citet{HjoPol93} are slightly extended to derive an asymptotic property for the estimator. Then, for the Kullback-Leibler divergence, we construct an asymptotically unbiased estimator by evaluating the asymptotic bias between the divergence and the log-likelihood into which the estimator is plugged. Moreover, we verify that this evaluation is the asymptotic bias in the strict sense; that is, the bias converges in mean to the evaluation. This sort of verification has usually been ignored in the literature (see, e.g., \citealt{KonKit08}). 

The rest of the paper is organized as follows. Section \ref{sec;Model} introduces the generalized linear model and the regularization method, and it describes some of the assumptions on our asymptotic theory. In Section \ref{sec;Asmptotics}, we discuss the asymptotic property of the estimator obtained from the regularization method, and in Section \ref{sec;IC}, we use it to evaluate the asymptotic bias, which is needed to derive the AIC. In Section \ref{sec;Moment}, we discuss the moment convergence of the estimator to show that the bias converges in mean to our evaluation. Section \ref{sec;Simulation} presents the results of simulation studies showing the validity of the proposed information criterion for several models, and Section \ref{sec;Discussion} gives concluding remarks and mentions future work. The proofs are relegated to the appendixes. 

\section{Setting and assumptions for asymptotics}
\label{sec;Model}
Let us consider a natural exponential family with a natural parameter $\bm{\theta}$ in $\Theta\;(\subset\mathbb{R}^{r})$ for an $r$-dimensional random variable $\bm{y}$, whose density is 
\begin{align*}
f(\bm{y};\bm{\theta})=\exp\left\{\bm{y}^{{\rm T}}\bm{\theta}-a(\bm{\theta})+b(\bm{y})\right\}
\end{align*}
with respect to a $\sigma$-finite measure. We assume that $\Theta$ is the natural parameter space; that is, $\bm{\theta}$ in $\Theta$ satisfies $0<\int\exp\{\bm{y}^{{\rm T}}\bm{\theta}+b(\bm{y})\}d\bm{y}<\infty$. Accordingly, all the derivatives of $a(\bm{\theta})$ and all the moments of $\bm{y}$ exist in the interior $\Theta^{{\rm int}}$ of $\Theta$, and, in particular, ${\rm E}[\bm{y}]=a'(\bm{\theta})$ and ${\rm V}[\bm{y}]=a''(\bm{\theta})$. For a function $c(\bm{\eta})$, we denote $\partial c(\bm{\eta})/\partial \bm{\eta}$ and $\partial^{2} c(\bm{\eta})/\partial \bm{\eta}\partial \bm{\eta}^{{\rm T}}$ by $c'(\bm{\eta})$ and $c''(\bm{\eta})$, respectively. We also assume that ${\rm V}[\bm{y}]=a''(\bm{\theta})$ is positive definite, and hence, $-\log f(\bm{y};\bm{\theta})$ is a strictly convex function with respect to $\bm{\theta}$.

Let $(\bm{y}_{i},\bm{X}_{i})$ be the $i$-th set of responses and regressors $(i=1,2,\ldots,n)$; we assume that $\bm{y}_{i}$ are independent $r$-dimensional random vectors and $\bm{X}_{i}$ in ${\cal X}\;(\subset\mathbb{R}^{r\times p})$ are $(r\times p)$-matrices of known constants. We will consider generalized linear models with natural link functions for such data (see \citealt{McCNel83}); that is, we will consider a class of density functions $\{f(\bm{y};\bm{X}\bm{\beta});\;\bm{\beta}\in{\cal B}\}$ for $\bm{y}_{i}$; thus, the log-likelihood function of $\bm{y}_{i}$ is given by
\begin{align*}
g_{i}(\bm{\beta})=\bm{y}_{i}^{{\rm T}}\bm{X}_{i}\bm{\beta}-a(\bm{X}_{i}\bm{\beta})+b(\bm{y}_{i}),
\end{align*}
where $\bm{\beta}$ is a $p$-dimensional coefficient vector and ${\cal B}\;(\subset\mathbb{R}^{p})$ is an open convex set. To develop an asymptotic theory for this model, we assume two conditions about the behavior of $\{\bm{X}_{i}\}$, as follows:
\begin{itemize}
\item[(C1)] ${\cal X}$ is a compact set with $\bm{X}\bm{\beta}\in\Theta^{{\rm int}}$ for all $\bm{X}\ (\in{\cal X})$ and $\bm{\beta}\ (\in{\cal B})$.
\item[(C2)] There exists an invariant distribution $\mu$ on ${\cal X}$. In particular, $n^{-1}\sum_{i=1}^{n}\bm{X}_{i}^{{\rm T}}a''(\bm{X}_{i}\bm{\beta})\bm{X}_{i}$ converges to a positive-definite matrix $\bm{J}(\bm{\beta})\equiv\int_{{\cal X}}\bm{X}^{{\rm T}}a''(\bm{X}\bm{\beta})\bm{X}\mu({\rm d}\bm{X})$.
\end{itemize}
In the above setting, we can prove the following lemma.
\begin{lem}
Let $\bm{\beta}^{*}$ be the true value of $\bm{\beta}$. Then, under conditions (C1) and (C2), we obtain the following:
\begin{itemize}
\item[(R1)] There exists a convex and differentiable function $h(\bm{\beta})$ such that $n^{-1}\sum_{i=1}^{n}\{g_{i}(\bm{\beta}^{*})-g_{i}(\bm{\beta})\} \stackrel{{\rm p}}{\to} h(\bm{\beta})$ for each $\bm{\beta}$.
\item[(R2)] $\bm{J}_{n}(\bm{\beta})\equiv -n^{-1}\sum_{i=1}^{n}g''_{i}(\bm{\beta})$ converges to $\bm{J}(\bm{\beta})$.
\item[(R3)] $\bm{s}_{n}\equiv n^{-1/2}\sum_{i=1}^{n}g'_{i}(\bm{\beta}^{*})\stackrel{{\rm d}}{\to}\bm{s}\sim {\rm N}(\bm{0},\bm{J}(\bm{\beta}^{*}))$.
\end{itemize}
\end{lem}
\noindent
See \citet{NinKaw14} for the proof. Note that we can explicitly write
\begin{align}
h(\bm{\beta})=\int_{\mathcal{X}}[a'(\bm{X}\bm{\beta}^{*})^{{\rm T}}\bm{X}(\bm{\beta}^{*}-\bm{\beta})-\{a(\bm{X}\bm{\beta}^{*})-a(\bm{X}\bm{\beta})\}]\mu({\rm d}\bm{X})
\label{defh}
\end{align}
since we assume (C2), and hence, we can prove its convexity and differentiability without using the techniques of convex analysis (\citealt{Roc70}).

Let us consider a non-concave penalized maximum likelihood estimator,
\begin{align}
\hat{\bm{\beta}}_{\lambda}=
\underset{\bm{\beta}\in{\cal B}}{{\rm argmin}}\left\{-\sum_{i=1}^{n}g_{i}(\bm{\beta})+n^{1/2}\sum_{j=1}^{p}p_{\lambda}(\beta_{j})\right\},
\label{eq;est}
\end{align}
where $\lambda\;(>0)$ is a tuning parameter and $p_{\lambda}(\beta_{j})$ is a penalty term with respect to $\beta_{j}$, which is not necessarily convex. Letting $q\in(0,1]$, we assume that $p_{\lambda}(\cdot)$ satisfies the following conditions; hereafter, we call it an $\ell_{q}$ type:
\begin{itemize}
\item[(C3)] $p_{\lambda}(\beta)$ is not differentiable only at the origin, symmetric with respect to $\beta=0$, and monotone non-decreasing with respect to $|\beta|$.
\item[(C4)] $\lim_{\beta\to0}p_{\lambda}(\beta)/|\beta|^{q}=\lambda$.
\end{itemize}
Such penalty terms for the Bridge, the SCAD, and the MCP are 
\begin{align*}
p_{\lambda}^{{\rm Bridge}}(\beta)
&=\lambda|\beta|^{q}, \\
p_{\lambda}^{{\rm SCAD}}(\beta)
&=\lambda|\beta|1_{\{|\beta|\leq(r+1)\lambda\}}-(|\beta|-\lambda)^2/(2r)1_{\{\lambda<|\beta|\leq (r+1)\lambda\}}+\lambda^{2}(1+r/2)1_{\{|\beta|>(r+1)\lambda\}},
\intertext{and}
p_{\lambda}^{{\rm MCP}}(\beta)
&=r\lambda^{2}/2-(r\lambda-|\beta|)^2/(2r)1_{\{|\beta|\leq r\lambda\}},
\end{align*}
where $0<q\leq 1$ and $r>1$. The Bridge penalty is the Lasso penalty itself when $q=1$, and it has the property that the derivative at the origin diverges when $0<q<1$. For the SCAD and MCP penalties, condition (C4) on the behavior in the neighborhood of the origin is satisfied by setting $q=1$, just like in the Lasso penalty. Thus, it is easy to imagine that a lot of penalties satisfy these conditions. Note that by using such penalties, several components of $\hat{\bm{\beta}}_{\lambda}$ tend to exactly 0 because of the non-differentiability at the origin. Also note that $p_{\lambda}(\cdot)$ is assumed not to depend on the subscript $j$ of the parameter for simplicity; this is not essential. While \citet{NinKaw14} put $n$ on the penalty term, we put $n^{1/2}$ on it in this study. From this, we can prove estimation consistency. Moreover, we can prove weak convergence of $n^{1/2}(\hat{\bm{\beta}}_{\lambda}-\bm{\beta}^{*})$, although the asymptotic distribution is not normal in general.

\section{Asymptotic behavior}
\label{sec;Asmptotics}
\subsection{Preparations}
Although the objective function in (\ref{eq;est}) is no longer convex because of the non-convexity of $p_{\lambda}(\cdot)$, the consistency of $\hat{\bm{\beta}}_{\lambda}$ can be derived by using a similar argument to the one in \citet{KniFu00}. First, the following lemma holds.
\begin{lem}\label{lem2}
$\hat{\bm{\beta}}_{\lambda}$ is a consistent estimator of $\bm{\beta}^{*}$, that is, $\hat{\bm{\beta}}_{\lambda}\stackrel{{\rm p}}{\to}\bm{\beta}^{*}$ under conditions (C1)--(C4).
\end{lem}
\noindent
This lemma is proved through uniform convergence of the random function,
\begin{align}
\mu_{n}(\bm{\beta})
=\frac{1}{n}\sum_{i=1}^{n}\{g_{i}(\bm{\beta}^{*})-g_{i}(\bm{\beta})\}
-\frac{1}{n^{1/2}}\sum_{j=1}^{p}\{p_{\lambda}(\beta_{j}^*)-p_{\lambda}(\beta_{j})\}.
\label{defmu}
\end{align}
The details are given in Section \ref{app;lem2}. Hereafter, we will denote $\bm{J}(\bm{\beta}^{*})$ by $\bm{J}$ so long as there is no confusion. In addition, we denote $\{j;\;\beta^{*}_{j}=0\}$ and $\{j;\;\beta^{*}_{j}\neq 0\}$ by ${\cal J}^{(1)}$ and ${\cal J}^{(2)}$, respectively. Moreover, the vector $(u_{j})_{j\in{\cal J}^{(k)}}$ and the matrix $(\bm{J}_{ij})_{i\in{\cal J}^{(k)},j\in{\cal J}^{(l)}}$ will be denoted by $\bm{u}^{(k)}$ and $\bm{J}^{(kl)}$, respectively, and we will sometimes express, for example, $\bm{u}$ as $(\bm{u}^{(1)},\bm{u}^{(2)})$.

To develop the asymptotic property of the penalized maximum likelihood estimator in (\ref{eq;est}), which will be used to derive an information criterion, we need to make a small generalization of the result in \citet{HjoPol93}, as follows:
\begin{lem}\label{lem3}
Suppose that $\eta_{n}(\bm{u})$ is a strictly convex random function that is approximated by $\tilde{\eta}_{n}(\bm{u})$. Let $\bm{u}^{\dagger}$ be a subvector of $\bm{u}$, and let $\phi(\bm{u})$ and $\psi(\bm{u}^{\dagger})$ be continuous functions such that $\phi_{n}(\bm{u})$ and $\psi_{n}(\bm{u}^{\dagger})$ converge to $\phi(\bm{u})$ and $\psi(\bm{u}^{\dagger})$ uniformly over $\bm{u}$ and $\bm{u}^{\dagger}$ in any compact set, respectively, and assume that $\phi(\bm{u})$ is convex and $\psi(\bm{0})=0$. In addition, for
\begin{align*}
\nu_{n}(\bm{u})=\eta_{n}(\bm{u})+\phi_{n}(\bm{u})+\psi_{n}(\bm{u}^{\dagger})\;\;\;\;\; {\rm and} \;\;\;\;\;
\tilde{\nu}_{n}(\bm{u})=\tilde{\eta}_{n}(\bm{u})+\phi(\bm{u})+\psi(\bm{u}^{\dagger}),
\end{align*}
let $\bm{u}_{n}$ and $\tilde{\bm{u}}_{n}$ be the argmin of $\nu_{n}(\bm{u})$ and $\tilde{\nu}_{n}(\bm{u})$, respectively, and assume that $\tilde{\bm{u}}_{n}$ is unique and $\tilde{\bm{u}}_{n}^{\dagger}=\bm{0}$. Then, for any $\varepsilon\;(>0)$, $\delta\;(>0)$ and $\xi\;(>\delta)$, there exists $\gamma\;(>0)$ such that
\begin{align}
{\rm P}(|\bm{u}_{n}-\tilde{\bm{u}}_{n}|\geq \delta)
\leq {\rm P}\left(2\Delta_{n}(\delta)+\varepsilon\geq \Upsilon_{n}(\delta)\right)
+{\rm P}(|\bm{u}_{n}-\tilde{\bm{u}}_{n}|\geq \xi)+{\rm P}(|\bm{u}_{n}^{\dagger}|>\gamma),
\label{eq;lem3_ineq}
\end{align}
where
\begin{align}
\Delta_{n}(\delta)=\sup_{|\bm{u}-\tilde{\bm{u}}_{n}|\leq \delta}|\nu_{n}(\bm{u})-\tilde{\nu}_{n}(\bm{u})|\;\;\;\;\; {\rm and} \;\;\;\;\;
\Upsilon_{n}(\delta)=\inf_{|\bm{u}-\tilde{\bm{u}}_{n}|=\delta}\tilde{\nu}_{n}(\bm{u})-\tilde{\nu}_{n}(\tilde{\bm{u}}_{n}).
\label{eq;lem3_check}
\end{align}
\end{lem}

\citet{HjoPol93} derived an inequality ${\rm P}(|\bm{u}_{n}-\tilde{\bm{u}}_{n}|\geq \delta)
\leq {\rm P}\left(2\Delta_{n}(\delta)\geq \Upsilon_{n}(\delta)\right)$; they assumed that $\nu_{n}(\bm{u})$ is convex. Although $\phi_{n}(\bm{u})+\psi_{n}(\bm{u}^{\dagger})$ is non-convex (hence $\nu_{n}(\bm{u})$ is too), we will use the fact that $\phi_{n}(\bm{u})+\psi_{n}(\bm{u}^{\dagger})$ converge to $\phi(\bm{u})+\psi(\bm{u}^{\dagger})$ over $\mathcal{U}\equiv\{\bm{u};\;|\bm{u}^{\dagger}|\leq \gamma,\;\delta\le|\bm{u}-\tilde{\bm{u}}_{n}|\le\xi\}$. In fact, if $n$ is sufficiently large, the inequality satisfied by the convex function is approximately satisfied for $\phi_{n}(\bm{u})$; that is, we have
\begin{align}
\left(1-\delta/l\right)\phi_{n}(\tilde{\bm{u}}_{n})+(\delta/l)\phi_{n}(\bm{u})-\phi_{n}(\tilde{\bm{u}}_{n}+\delta\bm{w}) 
> -\varepsilon/2
\label{prelem3_1_phi}
\end{align}
in $\mathcal{U}$.
Here, $\bm{w}$ is a unit vector such that $\bm{u}=\tilde{\bm{u}}_{n}+l\bm{w}$, and $l$ is in $[\delta,\xi]$, since $\delta\le|\bm{u}-\tilde{\bm{u}}_{n}|\le\xi$.
Moreover, if $\gamma$ is sufficiently small and $n$ is sufficiently large, since $\psi(\tilde{\bm{u}}_{n}^{\dagger})=0$, we have
\begin{align}
\left(1-\delta/l\right)\psi_{n}(\tilde{\bm{u}}_{n}^{\dagger})+(\delta/l)\psi_{n}(\bm{u}^{\dagger})-\psi_{n}(\tilde{\bm{u}}_{n}^{\dagger}+\delta\bm{w}^{\dagger}) 
> -\varepsilon/2
\label{prelem3_1_psi}
\end{align}
in $\mathcal{U}$. Hence, we can show that 
\begin{align}
{\rm P}(|\bm{u}_{n}^{\dagger}|\leq \gamma,\;\delta\leq |\bm{u}_{n}-\tilde{\bm{u}}_{n}|\leq \xi)
\leq {\rm P}(2\Delta_{n}(\delta)+\varepsilon\geq\Upsilon_{n}(\delta))
\label{prelem3_2}
\end{align}
in the same way as in \citet{HjoPol93}, from which we obtain the above lemma. See Section \ref{app;lem3} for the details. 

\subsection{Limiting distribution}
We use Lemma \ref{lem3} to derive the asymptotic property of the penalized maximum likelihood estimator in (\ref{eq;est}). Because the asymptotic property depends on the value of $q$, we will develop our argument by setting $0<q<1$. Furthermore, we will use $\tilde{q}=1/(2q)$ for the sake of simplicity.

Let us define a strictly convex random function,
\begin{align}
\eta_{n}(\bm{u}^{(1)},\bm{u}^{(2)})
=\sum_{i=1}^{n}\left\{g_{i}(\bm{\beta}^{*(1)},\bm{\beta}^{*(2)})
-g_{i}\left(\frac{\bm{u}^{(1)}}{n^{\tilde{q}}},\frac{\bm{u}^{(2)}}{n^{1/2}}+\bm{\beta}^{*(2)}\right)\right\}
\label{eq;eta_n}
\end{align}
and
\begin{align}
\tilde{\eta}_{n}(\bm{u}^{(1)},\bm{u}^{(2)})
=-\bm{u}^{(2){\rm T}}\bm{s}_{n}^{(2)}+\bm{u}^{(2){\rm T}}\bm{J}^{(22)}\bm{u}^{(2)}/2,
\label{eq;approx}
\end{align}
where $\bm{s}_{n}^{(2)}=n^{-1/2}\sum_{i=1}^{n}g'^{(2)}(\bm{\beta}^{*})$. By making a Taylor expansion around $(\bm{u}^{(1)},\bm{u}^{(2)})=(\bm{0},\bm{0})$, $\eta_{n}(\bm{u}^{(1)},\bm{u}^{(2)})$ can be expressed as
\begin{align*}
&-\sum_{i=1}^{n}\left\{\frac{1}{n^{\tilde{q}}}\bm{u}^{(1){\rm T}}g_{i}'^{(1)}(\bm{\beta}^{*})
+\frac{1}{n^{1/2}}\bm{u}^{(2){\rm T}}g_{i}'^{(2)}(\bm{\beta}^{*})\right\} \\
&-\sum_{i=1}^{n}\left\{
\frac{1}{2n^{2\tilde{q}}}\bm{u}^{(1){\rm T}}g_{i}''^{(11)}(\bm{\beta}^{*})\bm{u}^{(1)}
+\frac{1}{n^{\tilde{q}+1/2}}\bm{u}^{(1){\rm T}}g_{i}''^{(12)}(\bm{\beta}^{*})\bm{u}^{(2)}
+\frac{1}{2n}\bm{u}^{(2){\rm T}}g_{i}''^{(22)}(\bm{\beta}^{*})\bm{u}^{(2)}\right\}
\end{align*}
plus ${\rm o}_{{\rm p}}(1)$. Note that the term $-n^{-1}\sum_{i=1}^{n}\bm{u}^{(2){\rm T}}g_{i}''^{(22)}(\bm{\beta}^{*})\bm{u}^{(2)}$ converges to $\bm{u}^{(2){\rm T}}\bm{J}\bm{u}^{(2)}$ from (R2), and the terms including $\bm{u}^{(1)}$ reduce to ${\rm o}_{{\rm p}}(1)$. Accordingly, we see that $\eta_{n}(\bm{u}^{(1)},\bm{u}^{(2)})$ is asymptotically equivalent to $\tilde{\eta}_{n}(\bm{u}^{(1)},\bm{u}^{(2)})$. Next, letting $\bm{u}^{\dagger}$ be $\bm{u}^{(1)}$ and letting
\begin{align}
\phi_{n}(\bm{u})
=n^{1/2}\sum_{j\in{\cal J}^{(2)}}\left\{p_{\lambda}\left(\frac{u_{j}}{n^{1/2}}+\beta_{j}^{*}\right)-p_{\lambda}(\beta_{j}^{*})\right\}
\label{eq;phi_n}
\end{align}
and
\begin{align}
\psi_{n}(\bm{u}^{\dagger})=n^{1/2}\sum_{j\in{\cal J}^{(1)}}p_{\lambda}\left(\frac{u_{j}}{n^{\tilde{q}}}\right),
\label{eq;psi_n}
\end{align}
we can see from (C3) and (C4) that $\phi_{n}(\bm{u})$ and $\psi_{n}(\bm{u}^{\dagger})$ uniformly converge to a function,
\begin{align}
\phi(\bm{u})=\bm{u}^{(2){\rm T}}\bm{p}'^{(2)}_{\lambda}\;\;\;\;\;{\rm and}\;\;\;\;\;
\psi(\bm{u}^{\dagger})=\lambda\|\bm{u}^{(1)}\|_{q}^{q},
\label{eq;phi}
\end{align}
over $(\bm{u}^{(1)},\bm{u}^{(2)})$ in a compact set, respectively, where $\bm{p}'^{(2)}_{\lambda}=(p'_{\lambda}(\beta_{j}^{*}))_{j\in{\cal J}^{(2)}}$. In addition, letting $\nu_{n}(\bm{u}^{(1)},\bm{u}^{(2)})=\eta_{n}(\bm{u}^{(1)},\bm{u}^{(2)})+\phi_{n}(\bm{u})+\psi_{n}(\bm{u}^{\dagger})$ and $\tilde{\nu}_{n}(\bm{u}^{(1)},\bm{u}^{(2)})=\tilde{\eta}_{n}(\bm{u}^{(1)},\bm{u}^{(2)})+\phi(\bm{u})+\psi(\bm{u}^{\dagger})$, we see that the argmins of $\nu_{n}(\bm{u}^{(1)},\bm{u}^{(2)})$ and $\tilde{\nu}_{n}(\bm{u}^{(1)},\bm{u}^{(2)})$ are given by
\begin{align*}
(\bm{u}_{n}^{(1)},\bm{u}_{n}^{(2)})
=(n^{\tilde{q}}\hat{\bm{\beta}}^{(1)}_{\lambda},n^{1/2}(\hat{\bm{\beta}}^{(2)}_{\lambda}-\bm{\beta}^{*(2)}))
\;\;\;\;\;{\rm and}\;\;\;\;\;
 (\tilde{\bm{u}}_{n}^{(1)},\tilde{\bm{u}}_{n}^{(2)})
=(\bm{0},\bm{J}^{(22)-1}(\bm{s}_{n}^{(2)}-\bm{p}'^{(2)}_{\lambda})).
\end{align*}
Note that $\psi(\bm{u}^{\dagger})$ is not convex but satisfies that $\psi(\tilde{\bm{u}}_{n}^{(1)})=0$.
Using Lemma \ref{lem3} together with the above preliminaries, we find that, for any $\varepsilon\;(>0)$, $\delta\;(>0)$ and $\xi\;(>\delta)$, there exists $\gamma\;(>0)$ such that
\begin{align}
& {\rm P}(|(\bm{u}_{n}^{(1)},\bm{u}_{n}^{(2)}-\tilde{\bm{u}}_{n}^{(2)})|\geq \delta)
\nonumber \\
& \leq {\rm P}(2\Delta_{n}(\delta)+\varepsilon\geq \Upsilon_{n}(\delta)) 
+{\rm P}(|(\bm{u}_{n}^{(1)},\bm{u}_{n}^{(2)}-\tilde{\bm{u}}_{n}^{(2)})|\geq \xi)+{\rm P}(|\bm{u}_{n}^{(1)}|>\gamma),
\label{eq:lem3}
\end{align}
where $\Delta_{n}(\delta)$ and $\Upsilon_{n}(\delta)$ are the functions defined in (\ref{eq;lem3_check}). The triangle inequality, the convexity of $\eta_{n}(\bm{u}^{(1)},\bm{u}^{(2)})+\bm{u}^{(2){\rm T}}\bm{s}_{n}^{(2)}$ and the uniform convergence of $\phi_{n}(\bm{u})$ and $\psi_{n}(\bm{u}^{\dagger})$ imply
\begin{align}
\Delta_{n}(\delta)
\leq & \sup_{|(\bm{u}^{(1)},\bm{u}^{(2)}-\tilde{\bm{u}}_{n}^{(2)})|\leq \delta}|\eta_{n}(\bm{u}^{(1)},\bm{u}^{(2)})+\bm{u}^{(2){\rm T}}\bm{s}_{n}^{(2)}-\bm{u}^{(2){\rm T}}\bm{J}^{(22)}\bm{u}^{(2)}/2| \nonumber \\
& +\sup_{|(\bm{u}^{(1)},\bm{u}^{(2)}-\tilde{\bm{u}}_{n}^{(2)})|\leq \delta}|\phi_{n}(\bm{u})-\phi(\bm{u})| 
+\sup_{|(\bm{u}^{(1)},\bm{u}^{(2)}-\tilde{\bm{u}}_{n}^{(2)})|\leq \delta}|\psi_{n}(\bm{u}^{\dagger})-\psi(\bm{u}^{\dagger})| \nonumber \\
\stackrel{{\rm p}}{\to}&0.
\label{eq;Delta_conv}
\end{align}
Let $\rho\;(>0)$ be half the smallest eigenvalue of $\bm{J}^{(22)}$. Then, a simple calculation gives
\begin{align}
\Upsilon_{n}(\delta)
=\inf_{|(\bm{u}^{(1)},\bm{u}^{(2)}-\tilde{\bm{u}}_{n}^{(2)})|= \delta}\left\{\lambda\|\bm{u}^{(1)}\|_{q}^{q}+(\bm{u}^{(2)}-\tilde{\bm{u}}_{n}^{(2)})^{{\rm T}}\bm{J}^{(22)}(\bm{u}^{(2)}-\tilde{\bm{u}}_{n}^{(2)})/2\right\}
\geq\min\{\lambda\delta^{q},\rho\delta^{2}\}.
\label{eq:Upsilon_min}
\end{align}
From (\ref{eq;Delta_conv}) and (\ref{eq:Upsilon_min}), by considering a sufficiently small $\varepsilon$ and a sufficiently large $n$, the first term on the right-hand side in (\ref{eq:lem3}) can be made arbitrarily small. In addition, we can generalize the result in \citet{Rad05} with respect to the model and the penalty term; thus, for any $\gamma\;(>0)$, we have
\begin{align}
{\rm P}(|\bm{u}_{n}^{(1)}|\leq \gamma)\to1\;\;\;\;\;{\rm and}\;\;\;\;\;
|\bm{u}_{n}-\tilde{\bm{u}}_{n}|={\rm O}_{{\rm p}}(1).
\label{eq;Rad}
\end{align}
See Section \ref{app;Rad} for the proof of (\ref{eq;Rad}). From this, by considering a sufficiently large $\xi$ and a sufficiently large $n$, the second and third terms on the right-hand side in (\ref{eq:lem3}) can be made arbitrarily small. Thus, we conclude that
\begin{align*}
\bm{u}_{n}^{(1)}={\rm o}_{{\rm p}}(1)\;\;\;\;\;{\rm and}\;\;\;\;\;\bm{u}_{n}^{(2)}=\tilde{\bm{u}}_{n}^{(2)}+{\rm o}_{{\rm p}}(1).
\end{align*}
\begin{thm}\label{thm1}
Let $\bm{p}'^{(2)}_{\lambda}=(p'_{\lambda}(\beta_{j}^{*}))_{j\in{\cal J}^{(2)}},\; \bm{J}^{(1|2)}=\bm{J}^{(11)}-\bm{J}^{(12)}\bm{J}^{(22)-1}\bm{J}^{(21)},\;\bm{\tau}_{\lambda}(\bm{s}_{n})=\bm{s}_{n}^{(1)}-\bm{J}^{(12)}\bm{J}^{(22)-1}(\bm{s}_{n}^{(2)}-\bm{p}'^{(2)}_{\lambda})$ and
\begin{align}
\hat{\bm{u}}_{n}^{(1)}
=\underset{\bm{u}^{(1)}}{{\rm argmin}}\left\{\bm{u}^{(1){\rm T}}\bm{J}^{(1|2)}\bm{u}^{(1)}/2-\bm{u}^{(1){\rm T}}\bm{\tau}_{\lambda}(\bm{s}_{n})+\lambda\|\bm{u}^{(1)}\|_{1}\right\}.
\label{eq;u1}
\end{align}
Under conditions (C1)--(C4), we have
\begin{align*}
n^{1/(2q)}\hat{\bm{\beta}}_{\lambda}^{(1)}={\rm o}_{{\rm p}}(1)
\;\;\;\;\;{\rm and}\;\;\;\;\;
n^{1/2}(\hat{\bm{\beta}}_{\lambda}^{(2)}-\bm{\beta}^{*(2)})
=\bm{J}^{(22)-1}(\bm{s}_{n}^{(2)}-\bm{p}'^{(2)}_{\lambda})+{\rm o}_{{\rm p}}(1)
\end{align*}
when $0<q<1$, and we have
\begin{align}
n^{1/2}\hat{\bm{\beta}}_{\lambda}^{(1)}
=\hat{\bm{u}}_{n}^{(1)}+{\rm o}_{{\rm p}}(1)
\label{eq;a1-1}
\end{align}
and
\begin{align}
n^{1/2}(\hat{\bm{\beta}}_{\lambda}^{(2)}-\bm{\beta}^{*(2)})
=-\bm{J}^{(22)-1}\bm{J}^{(21)}\hat{\bm{u}}_{n}^{(1)}+\bm{J}^{(22)-1}(\bm{s}_{n}^{(2)}-\bm{p}'^{(2)}_{\lambda})+{\rm o}_{{\rm p}}(1)
\label{eq;a1-2}
\end{align}
when $q=1$.
\end{thm}
\noindent
We can obtain the result for the case of $q=1$ in almost the same way as in the case of $0<q<1$ (see Section \ref{app;thm1} for details). From Theorem \ref{thm1}, the estimator $\hat{\bm{\beta}}_{\lambda}$ in (\ref{eq;est}) is shown to converge in distribution to some function of a Gaussian distributed random variable. When $0<q<1$, we immediately see that it is 0 or the Gaussian distributed random variable itself, and this simple fact is useful for deriving an information criterion explicitly and reducing the computational cost of model selection. On the other hand, when $q=1$, we can prove weak convergence, since the convex objective function in (\ref{eq;u1}) converges uniformly from the convexity lemma in \citet{HjoPol93}.
\begin{coro}
Let $\bm{s}^{(2)}$ be a Gaussian distributed random variable with mean $\bm{0}$ and covariance matrix $\bm{J}^{(22)}$ and
\begin{align}
\hat{\bm{u}}^{(1)}
=\underset{\bm{u}^{(1)}}{{\rm argmin}}\left\{\bm{u}^{(1){\rm T}}\bm{J}^{(1|2)}\bm{u}^{(1)}/2-\bm{u}^{(1){\rm T}}\bm{\tau}_{\lambda}(\bm{s})+\lambda\|\bm{u}^{(1)}\|_{1}\right\}.
\label{eq;asy}
\end{align}
Then, under the same conditions as in Theorem \ref{thm1}, we have
\begin{align*}
n^{1/(2q)}\hat{\bm{\beta}}_{\lambda}^{(1)}\stackrel{{\rm d}}{\to}\bm{0}
\;\;\;\;\;{\rm and}\;\;\;\;\;
n^{1/2}(\hat{\bm{\beta}}_{\lambda}^{(2)}-\bm{\beta}^{*(2)})
\stackrel{{\rm d}}{\to}\bm{J}^{(22)-1}(\bm{s}^{(2)}-\bm{p}'^{(2)}_{\lambda})
\end{align*}
when $0<q<1$, and we have
\begin{align*}
n^{1/2}\hat{\bm{\beta}}_{\lambda}^{(1)}
\stackrel{{\rm d}}{\to}\hat{\bm{u}}^{(1)}
\;\;\;\;\;{\rm and}\;\;\;\;\;
n^{1/2}(\hat{\bm{\beta}}_{\lambda}^{(2)}-\bm{\beta}^{*(2)})
\stackrel{{\rm d}}{\to}-\bm{J}^{(22)-1}\bm{J}^{(21)}\hat{\bm{u}}^{(1)}+\bm{J}^{(22)-1}(\bm{s}^{(2)}-\bm{p}'^{(2)}_{\lambda})
\end{align*}
when $q=1$.
\label{coro1}
\end{coro}
In the case of $q=1$, we still need to solve the minimization problem in (\ref{eq;asy}) for evaluating the AIC, but this is easy because the objective function is convex with respect to $\bm{u}^{(1)}$, so we can use existing convex optimization techniques. It is known that the proximal gradient method (\citealt{Roc76,BecTeb09}) is effective for solving such a minimization problem when the objective function is the sum of a differentiable function and a non-differentiable function. We will use, however, the coordinate descent method (\citealt{MazFriHas11}) because the objective function can be minimized explicitly for each variable. Actually, when we fix all the elements of $\hat{\bm{u}}$ except for the $j$-th one, $\hat{u}_{j}^{(1)}$ is given by
\begin{align*}
\hat{u}_{j}^{(1)}=\frac{1}{\bm{J}_{jj}^{(1|2)}}{\rm sgn}\left(\tau_{j}-\sum_{k\neq j}\bm{J}_{jk}^{(1|2)}\hat{u}_{k}^{(1)}\right)\max\left\{\left|\tau_{j}-\sum_{k\neq j}\bm{J}_{jk}^{(1|2)}\hat{u}_{k}^{(1)}\right|-\lambda,0\right\}.
\end{align*}
Then, for the $(t+1)$-th step in the algorithm, we have only to update $u_{j}^{(t)}$ as follows:
\begin{align*}
u^{(t+1)}_{j}
=\underset{u}{{\rm argmin}} \;h(u_{1}^{(t+1)},u_{2}^{(t+1)},\ldots,u_{j-1}^{(t+1)},u,u_{j+1}^{(t)},u_{j+2}^{(t)},\ldots,u_{|{\cal J}^{(1)}|}^{(t)}),
\end{align*}
for $j=1,2,\ldots,|{\cal J}^{(1)}|$, and we repeat this update until $|\bm{u}^{(t+1)}-\bm{u}^{(t)}|$ converges. Note that the optimal value $\hat{u}_{j}^{(1)}$ satisfies $\hat{u}_{j}^{(1)}=0$ if $|(\bm{J}^{(1|2)}\hat{\bm{u}}+\bm{\tau}_{\lambda}(\bm{s}))_{j}|\leq \lambda$ and $(\bm{J}^{(1|2)}\hat{\bm{u}}+\bm{\tau}_{\lambda}(\bm{s}))_{j}=-\lambda{\rm sgn}(\hat{u}_{j}^{(1)})$ otherwise.

\section{Information criterion}
\label{sec;IC}
From the perspective of prediction, model selection using the AIC aims to minimize twice the Kullback-Leibler divergence (\citealt{KulLei51}) between the true distribution and the estimated distribution, 
\begin{align*}
2\tilde{{\rm E}}\left[\sum_{i=1}^{n}\tilde{g}_{i}(\bm{\beta}^{*})\right]-2\tilde{{\rm E}}\left[\sum_{i=1}^{n}\tilde{g}_{i}(\hat{\bm{\beta}}_{\lambda})\right],
\end{align*}
where $(\tilde{\bm{y}}_{1},\tilde{\bm{y}}_{2},\ldots,\tilde{\bm{y}}_{n})$ is a copy of $(\bm{y}_{1},\bm{y}_{2},\ldots,\bm{y}_{n})$; in other words, $(\tilde{\bm{y}}_{1},\tilde{\bm{y}}_{2},\ldots,\tilde{\bm{y}}_{n})$ has the same distribution as $(\bm{y}_{1},\bm{y}_{2},\ldots,\bm{y}_{n})$ and is independent of $(\bm{y}_{1},\bm{y}_{2},\ldots,\bm{y}_{n})$. In addition, $\tilde{g}_{i}(\bm{\beta})$ and $\tilde{{\rm E}}$ denote a log-likelihood function based on $\tilde{\bm{y}}_{i}$, that is, $\log f(\tilde{\bm{y}}_{i};\bm{X}_{i}\bm{\beta})$, and the expectation with respect to only $(\tilde{\bm{y}}_{1},\tilde{\bm{y}}_{2},\ldots,\tilde{\bm{y}}_{n})$, respectively. Because the first term is a constant, i.e., it does not depend on the model selection, we only need to consider the second term, and then the AIC is defined as an asymptotically biased estimator for it (\citealt{aka73}). A simple estimator of the second term in our setting is $-2\sum_{i=1}^{n}g_{i}(\hat{\bm{\beta}}_{\lambda})$, but it underestimates the second term. Consequently, we will minimize the bias correction, 
\begin{align}
-2\sum_{i=1}^{n}g_{i}(\hat{\bm{\beta}}_{\lambda})
+2{\rm E}\left[\sum_{i=1}^{n}g_{i}(\hat{\bm{\beta}}_{\lambda})-\tilde{{\rm E}}\left[\sum_{i=1}^{n}\tilde{g}_{i}(\hat{\bm{\beta}}_{\lambda})\right]\right],
\label{eq;ic}
\end{align}
in AIC-type information criteria (see \citealt{KonKit08}). Because the expectation in (\ref{eq;ic}), i.e., the bias term, depends on the true distribution, it cannot be explicitly given in general; thus, we will evaluate it asymptotically in the same way as was done for the AIC.

For the Lasso, \citet{EfrHasJohTib04} and \citet{ZouHasTib07} developed the $C_{p}$-type information criterion as an unbiased estimator of the prediction squared error in a Gaussian linear regression setting, in other words, a finite correction of the AIC (\citealt{Sug78}) in a Gaussian linear setting with a known variance. For the Lasso estimator $\hat{\bm{\beta}}_{\lambda}=(\hat{\beta}_{\lambda,1},\ldots,\hat{\beta}_{\lambda,p})$, it can be expressed as
\begin{align*}
\sum_{i=1}^{n}\{(\bm{y}_{i}-\bm{X}_{i}\hat{\bm{\beta}}_{\lambda})^{{\rm T}}{\rm V}[\bm{y}_{i}]^{-1}(\bm{y}_{i}-\bm{X}_{i}\hat{\bm{\beta}}_{\lambda})+\log|2\pi{\rm V}[\bm{y}_{i}]|\}
+2|\{ j;\; \hat{\beta}_{\lambda,j}\neq 0\}|,
\end{align*}
where the index set $\{j ;\;\hat{\beta}_{\lambda,j}\neq 0\}$ is called an active set. Unfortunately, since Stein's unbiased risk estimation theory (\citealt{Ste81}) was used for deriving this criterion, it was difficult to extend this result to other models. In that situation, \citet{NinKaw14} relied on statistical asymptotic theory and extended the result to generalized linear models based on the asymptotic distribution of the Lasso estimator. The Lasso estimator in their paper is defined by
\begin{align*}
\hat{\bm{\beta}}_{\lambda}
=\underset{\bm{\beta}\in{\cal B}}{{\rm argmin}}\left\{-\sum_{i=1}^{n}g_{i}(\bm{\beta})+n\lambda\|\bm{\beta}\|_{1}\right\},
\end{align*}
but, as was mentioned in the previous section, estimation consistency is not assured because the order of the penalty term is ${\rm O}(n)$. In this study, we derive an information criterion in a setting that estimation consistency holds as in Lemma \ref{lem2} for not only the Lasso but also the non-concave penalized likelihood method.

The bias term in (\ref{eq;ic}) can be rewritten as the expectation of
\begin{align}
\sum_{i=1}^{n}\{g_{i}(\hat{\bm{\beta}}_{\lambda})-g_{i}(\bm{\beta}^{*})\}
-\sum_{i=1}^{n}\{\tilde{g}_{i}(\hat{\bm{\beta}}_{\lambda})-\tilde{g}_{i}(\bm{\beta}^{*})\},
\label{eq;bias}
\end{align}
so we can derive an AIC by evaluating ${\rm E}[z^{{\rm limit}}]$, where $z^{{\rm limit}}$ is the limit to which (\ref{eq;bias}) converges in distribution. We call ${\rm E}[z^{{\rm limit}}]$ an asymptotic bias. Here, we will develop an argument by setting $0<q<1$.

Using Taylor's theorem, the first term in (\ref{eq;bias}) can be expressed as
\begin{align}
(\hat{\bm{\beta}}_{\lambda}-\bm{\beta}^{*})^{{\rm T}}\sum_{i=1}^{n}g'_{i}(\bm{\beta}^{*})
+(\hat{\bm{\beta}}_{\lambda}-\bm{\beta}^{*})^{{\rm T}}\sum_{i=1}^{n}g''_{i}(\bm{\beta}^{\dagger})(\hat{\bm{\beta}}_{\lambda}-\bm{\beta}^{*})/2,
\label{eq;taylor}
\end{align}
where $\bm{\beta}^{\dagger}$ is a vector on the segment from $\hat{\bm{\beta}}_{\lambda}$ to $\bm{\beta}^{*}$. Note that $-n^{-1}\sum_{i=1}^{n}g''_{i}(\bm{\beta}^{\dagger})$ converges in probability to $\bm{J}$ from (R2) and Lemma \ref{lem2}. Now we apply Theorem \ref{thm1}. First, the terms including $\hat{\bm{\beta}}_{\lambda}^{(1)}$ reduce to ${\rm o}_{{\rm p}}(1)$ because $n^{1/(2q)}\hat{\bm{\beta}}_{\lambda}^{(1)}={\rm o}_{{\rm p}}(1)$. Moreover, $n^{1/2}(\hat{\bm{\beta}}^{(2)}_{\lambda}-\bm{\beta}^{*})$ is asymptotically equivalent to $\bm{J}^{(22)-1}(\bm{s}_{n}^{(2)}-\bm{p}'^{(2)}_{\lambda})$. Thus, (\ref{eq;taylor}) can be expressed as
\begin{align*}
\bm{s}_{n}^{(2){\rm T}}\bm{J}^{(22)-1}(\bm{s}_{n}^{(2)}-\bm{p}'^{(2)}_{\lambda})-(\bm{s}_{n}^{(2)}-\bm{p}'^{(2)}_{\lambda})^{{\rm T}}\bm{J}^{(22)-1}(\bm{s}_{n}^{(2)}-\bm{p}'^{(2)}_{\lambda})/2+{\rm o}_{{\rm p}}(1),
\end{align*}
and we see that this converges in distribution to
\begin{align*}
\bm{s}^{(2){\rm T}}\bm{J}^{(22)-1}(\bm{s}^{(2)}-\bm{p}'^{(2)}_{\lambda})-(\bm{s}^{(2)}-\bm{p}'^{(2)}_{\lambda})^{{\rm T}}\bm{J}^{(22)-1}(\bm{s}^{(2)}-\bm{p}'^{(2)}_{\lambda})/2
\end{align*}
from (R3). Similarly, the second term in (\ref{eq;bias}) can be expressed as using Taylor's theorem
\begin{align}
(\hat{\bm{\beta}}_{\lambda}-\bm{\beta}^{*})^{{\rm T}}\sum_{i=1}^{n}\tilde{g}'_{i}(\bm{\beta}^{*})
+(\hat{\bm{\beta}}_{\lambda}-\bm{\beta}^{*})^{{\rm T}}\sum_{i=1}^{n}\tilde{g}''_{i}(\bm{\beta}^{\ddagger})(\hat{\bm{\beta}}_{\lambda}-\bm{\beta}^{*})/2,
\label{eq;taylor_copy}
\end{align}
where $\bm{\beta}^{\ddagger}$ is a vector on the segment from $\hat{\bm{\beta}}_{\lambda}$ to $\bm{\beta}^{*}$, and by applying Theorem \ref{thm1} and (R3), we see that this converges in distribution to
\begin{align*}
\tilde{\bm{s}}^{(2){\rm T}}\bm{J}^{(22)-1}(\bm{s}^{(2)}-\bm{p}'^{(2)}_{\lambda})-(\bm{s}^{(2)}-\bm{p}'^{(2)}_{\lambda})^{{\rm T}}\bm{J}^{(22)-1}(\bm{s}^{(2)}-\bm{p}'^{(2)}_{\lambda})/2,
\end{align*}
where $\tilde{\bm{s}}^{(2)}$ is a copy of $\bm{s}^{(2)}$. Hence, we have
\begin{align*}
z^{{\rm limit}}
=\bm{s}^{(2){\rm T}}\bm{J}^{(22)-1}(\bm{s}^{(2)}-\bm{p}'^{(2)}_{\lambda})-\tilde{\bm{s}}^{(2){\rm T}}\bm{J}^{(22)-1}(\bm{s}^{(2)}-\bm{p}'^{(2)}_{\lambda}).
\end{align*}
Because $\bm{s}^{(2)}$ and $\tilde{\bm{s}}^{(2)}$ are independently distributed according to ${\rm N}(\bm{0},\bm{J}^{(22)})$, the asymptotic bias reduces to
\begin{align*}
{\rm E}[z^{{\rm limit}}]={\rm E}[\bm{s}^{(2){\rm T}}\bm{J}^{(22)-1}(\bm{s}^{(2)}-\bm{p}'^{(2)}_{\lambda})],
\end{align*}
and we obtain the following theorem.
\begin{thm}\label{thm2}
Under the same conditions as in Theorem \ref{thm1}, we have
\begin{align*}
{\rm E}[z^{{\rm limit}}]=|{\cal J}^{(2)}|
\end{align*}
when $0<q<1$, and we have
\begin{align}
{\rm E}[z^{{\rm limit}}]
=|{\cal J}^{(2)}|+K
\label{eq;a1-3}
\end{align}
when $q=1$, where $K={\rm E}\left[\hat{\bm{u}}^{(1){\rm T}}\bm{s}^{(1|2)}\right],\;\bm{s}^{(1|2)}=\bm{s}^{(1)}-\bm{J}^{(12)}\bm{J}^{(22)-1}\bm{s}^{(2)}$, and $\hat{\bm{u}}^{(1)}$ is the random vector defined in (\ref{eq;asy}).
\end{thm}
\noindent
We can obtain the result in the case of $q=1$ in almost the same way as in the case of $0<q<1$ (see Section \ref{app;thm2} for details). Because the asymptotic bias derived in Theorem \ref{thm2} depends on an unknown value $\bm{\beta}^{*}$, we need to evaluate it. Here, we use the fact that $\hat{\bm{\beta}}_{\lambda}$ is a consistent estimator of $\bm{\beta}^{*}$ from Lemma \ref{lem2} and that $\bm{J}_{n}(\hat{\bm{\beta}}_{\lambda})=n^{-1}\sum_{i=1}^{n}\bm{X}^{{\rm T}}a''(\bm{X}\hat{\bm{\beta}}_{\lambda})\bm{X}$ converges in probability to $\bm{J}$. Concretely speaking, we replace ${\cal J}^{(2)}$ by the active set $\hat{{\cal J}}^{(2)}=\{j;\;\hat{\beta}_{\lambda,j}\neq 0\}$ and $K$ by its empirical mean $\hat{K}$ obtained by generating samples from ${\rm N}(\bm{0},\bm{J}_{n}(\hat{\bm{\beta}}_{\lambda}))$. As a result, we propose the following index as an AIC for the non-concave penalized maximum likelihood method:
\begin{align}
{\rm AIC}_{\lambda}^{\ell_{q}\mathchar`-{\rm type}}
 =\left\{ \begin{array}{lc}
 \displaystyle{-2\sum_{i=1}^{n}g_{i}(\hat{\bm{\beta}}_{\lambda})+2| \hat{{\cal J}}^{(2)}|}&(0<q<1) \\
 \displaystyle{-2\sum_{i=1}^{n}g_{i}(\hat{\bm{\beta}}_{\lambda})+2| \hat{{\cal J}}^{(2)}|+2\hat{K}}&(q=1)
\end{array} \right..
\label{eq;aic}
\end{align}
When $0<q<1$, we can see that the bias term of the information criterion in \citet{EfrHasJohTib04} or \citet{ZouHasTib07} can be used not only for Gaussian linear regression settings but also for generalized linear settings. Thus, by minimizing the AIC in (\ref{eq;aic}), we can obtain the optimal value of the tuning parameter $\lambda$.

\section{Moment Convergence}
\label{sec;Moment}
By adding trivial conditions, we can verify that convergence holds in mean for the asymptotic bias in Theorem \ref{thm2}; that is, the second term in (\ref{eq;ic}) converges to $|{\cal J}^{(2)}|$ when $0<q<1$ and $|{\cal J}^{(2)}|+K$ when $q=1$. Note that this sort of verification is usually ignored in the literature (see, e.g., \citealt{KonKit08}). 

To deal with the cases of $0<q<1$ and $q=1$ simultaneously, let us denote $\sum_{i=1}^n\{g_i(\bm{\beta}^*)-g_i(n^{-1/2}\bm{u}+\bm{\beta}^*)\}-n^{1/2}\sum_{j=1}^p\{p_{\lambda}(\beta_j^*)-p_{\lambda}(n^{-1/2}u_j+\beta_j^*)\}$ by $\nu_n(\bm{u})$ also for $0<q<1$ in this section and the weak limit of $\bm{u}_n=\argmin_{\bm{u}}\nu_n(\bm{u})$ by
\begin{align*}
\tilde{\bm{u}}=(\tilde{\bm{u}}^{(1)},\tilde{\bm{u}}^{(2)})
=(\hat{\bm{u}}^{(1)}1_{\{q=1\}},-\bm{J}^{(22)-1}\bm{J}^{(21)}\hat{\bm{u}}^{(1)}1_{\{q=1\}}+\bm{J}^{(22)-1}(\bm{s}^{(2)}-\bm{p}'^{(2)}_{\lambda}))
\end{align*}
which is given in Corollary \ref{coro1}. 

First, we state the result of applying the theorem in \cite{Yos11} to our problem, which gives sufficient conditions for a polynomial-type large deviation inequality with respect to $\bm{u}_n$. Note that the theorem in \cite{Yos11} also plays an essential role in \cite{MasShi14}.
In this section, we assume that ${\cal B}$ is a precompact set.
Letting $\alpha\in(0,1)$, $L>2$ and $\omega_n(\bm{u})=\nu_n(\bm{u})-n^{1/2}\sum_{j\in{\cal J}^{(1)}}p_{\lambda}(n^{-1/2}u_{j})+\bm{u}^{\rm T}\bm{s}_n-\bm{u}^{\rm T}\bm{J}\bm{u}/2$, the sufficient conditions can be written as follows:
\begin{itemize}
\item[(A1)] $\exists \chi_1=\chi_1(\bm{\beta}^*)>0,\ \exists \chi_2=\chi_2(\bm{\beta}^*)>0,\ \forall\bm{\beta}\in\mathcal{B}$,
\begin{align}
h(\bm{\beta})\geq\chi_1\left| \bm{\beta}-\bm{\beta}^*\right|^{\chi_2}. \nonumber
\end{align}
\item[(A2)] $\exists \gamma_{1}>0,\ \exists c_{1}>0$,
\begin{align*}
\sup_{r>0}\sup_{n>0}r^{L}{\rm P}\left(\sup_{\bm{u}\in U_{n}(r)}\frac{|\omega_{n}(\bm{u})|}{1+| \bm{u}|^{2}}\geq r^{-\gamma_{1}}\right)\leq c_{1},
\end{align*}
where $U_{n}(r)=\{\bm{u}\in\mathbb{R}^{p};\ r\leq |\bm{u}|\leq n^{(1-\alpha)/2}\}$. 
\item[(A3)] $\exists \gamma_{2}\in[0,1/2),\ \exists c_{2}\in(\alpha\chi_2,1-2\gamma_2)$, 
\begin{align*}
\sup_{n>0}{\rm E}[\left|\bm{s}_n\right|^{N_{1}}]<\infty
\;\;\;\;\; {\rm and} \;\;\;\;\;\sup_{n>0}{\rm E}\left[\sup_{\bm{\beta}\in\mathcal{B}}\{n^{1/2-\gamma_{2}}\left|\mu_n(\bm{\beta})-h(\bm{\beta})\right|\}^{N_{2}}\right]<\infty
\end{align*}
where $N_{1}=L(1-\gamma_{1})^{-1}$, $N_{2}=L(1-2\gamma_2-c_{2})^{-1}$, and $\mu_n(\bm{\beta})$ is the random function defined in (\ref{defmu}). 
\end{itemize}

\begin{thm}[\citealt{Yos11}]
If there exists $\alpha\ (\in(0,1))$ such that (A1)--(A3) hold, we have
\begin{align}
\sup_{r>0}\sup_{n>0}r^{L}{\rm P}\left(\sup_{|\bm{u}|\geq r}
\left\{-\nu_{n}(\bm{u})\right\}\geq 0\right)<\infty. 
\label{eq;pldi}
\end{align}
\label{thm_NY2011}
\end{thm}

The definition of $\omega_{n}(\bm{u})$ may seem somewhat strange, but this can be justified from the non-negativity of $p_{\lambda}(\cdot)$.
In fact, we see that
\begin{align*}
{\rm P}\left(\sup_{|\bm{u}|\geq r}
\left\{-\nu_{n}(\bm{u})\right\}\geq 0\right)
\leq {\rm P}\left(\sup_{|\bm{u}|\geq r}
\left\{-\nu_{n}(\bm{u})+n^{1/2}\sum_{j\in{\cal J}^{(1)}}p_{\lambda}\left(\frac{u_{j}}{n^{1/2}}\right)\right\}\geq 0\right).
\end{align*}
Therefore, to obtain (\ref{eq;pldi}), it suffices to establish a polynomial-type large deviation inequality for a random function $-\nu_{n}(\bm{u})+n^{1/2}\sum_{j\in{\cal J}^{(1)}}p_{\lambda}(n^{-1/2}u_{j})$ instead of $-\nu_{n}(\bm{u})$.

We can easily obtain from (\ref{eq;pldi}) that 
\begin{align*}
\sup_{r>0}\sup_{n>0}r^{L}{\rm P}\left(\left|\bm{u}_{n}\right|\geq r\right)<\infty. 
\end{align*}
Moreover, considering the weak convergence of $\bm{u}_{n}$ to $\tilde{\bm{u}}$, we have
\begin{align}
{\rm E}\left[f_L(\bm{u}_{n})\right]\rightarrow {\rm E}\left[f_L(\tilde{\bm{u}})\right]
\label{eq;w_moment}
\end{align}
for every polynomial growth function $f_L:\mathbb{R}^{p}\rightarrow \mathbb{R}$ whose order is less than $L$. 

The sufficient conditions (A1)--(A3) can not be derived from only (C1)--(C4); we require additional trivial conditions:
\begin{itemize}
\item[(C5)] The eigenvalues of $\bm{J}(\bm{\beta})$ are uniformly bounded away from 0 and infinity over $\bm{\beta}\in\mathcal{B}$.
\item[(C6)] There exists $\delta_{1}\ (\in (0,1))$ such that
\begin{align*}
\sup_{n>0}\left\{n^{\delta_{1}}\left|\frac{1}{n}\sum_{i=1}^{n}g_{i}''(\bm{\beta}^*)+\bm{J}\right|\right\}<\infty. 
\end{align*}
\item[(C7)] There exists $\delta_{2}\ (\in (0,1))$ such that 
\begin{align*}
\sup_{n>0}{\rm E}\left[\left\{n^{\delta_{2}}\left|\frac{1}{n}\sum_{i=1}^{n}y_{i}^{\rm T}\bm{X}_{i}-\int_{\mathcal{X}}a'(\bm{X}\bm{\beta}^*)^{\rm T}\bm{X}\mu({\rm d}\bm{X})\right|\right\}^{k}\right]<\infty
\end{align*}
for all $k\ (\in\mathbb{N})$ and 
\begin{align*}
\sup_{n>0}\sup_{\bm{\beta}\in\mathcal{B}}\left\{n^{\delta_{2}}\left|\frac{1}{n}\sum_{i=1}^{n}a(\bm{X}_{i}\bm{\beta})-\int_{\mathcal{X}}a(\bm{X}\bm{\beta})\mu({\rm d}\bm{X})\right|\right\}<\infty.
\end{align*}
\end{itemize}

Letting $\alpha\in (0,\min\{2\delta_1,\delta_2,1/2\})$, we will check the sufficient conditions.

First, it can be easily seen from (C5) that (A1) holds by setting $\chi_1$ to the infimum of the smallest eigenvalue of $\bm{J}(\bm{\beta})$ over $\bm{\beta}\in\mathcal{B}$ and $\chi_2=2$, as we obtain
\begin{align*}
h(\bm{\beta})=\int_{\mathcal{X}}\left\{(\bm{\beta}-\bm{\beta}^*)^{\rm T}\bm{X}^{\rm T}a''(\bm{X}\tilde{\bm{\beta}})\bm{X}(\bm{\beta}-\bm{\beta}^*)\right\}\mu({\rm d}\bm{X})
\end{align*}
from using Taylor's theorem for $h(\bm{\beta})$ in (\ref{defh}), where $\tilde{\bm{\beta}}$ is a vector between $\bm{\beta}$ and $\bm{\beta}^*$. 

Next, let us consider (A2). Using Taylor's theorem, $\omega_n(\bm{u})$ can be written as
\begin{align*}
-\bm{u}^{\rm T}\int_{0}^{1}(1-s)
\left\{\frac{1}{n}\sum_{i=1}^{n}g_{i}''\left(\bm{\beta}^*+\frac{\bm{u}s}{n^{1/2}}\right)+\bm{J}\right\}
{\rm d}s\bm{u}
+n^{1/2}\sum_{j\in\mathcal{J}^{(2)}}\left\{p_{\lambda}\left(\beta_{j}^*+\frac{u_{j}}{n^{1/2}}\right)-p_{\lambda}(\beta_{j}^*)\right\}. 
\end{align*}
Using Taylor's theorem again for $g_{i}''(\bm{\beta}^*+n^{-1/2}\bm{u}s)$ and (C3), we get
\begin{align}
\frac{|\omega_{n}(\bm{u})|}{1+|\bm{u}|^{2}}\lesssim&\frac{|\bm{u}|^{2}}{1+|\bm{u}|^{2}}\left|\frac{1}{n}\sum_{i=1}^{n}g_{i}''(\bm{\beta}^*)+\bm{J}\right|
\nonumber \\
&+\frac{|\bm{u}|^{2}}{1+|\bm{u}|^{2}}\frac{|\bm{u}|}{n^{1/2}}\int_{0}^{1}\int_{0}^{1}\left|\frac{1}{n}\sum_{i=1}^{n}g_{i}'''\left(\bm{\beta}^*+\frac{\bm{u}st}{n^{1/2}}\right)\right|{\rm d}t{\rm d}s +\frac{|\bm{u}|}{1+|\bm{u}|^{2}},
\label{eq:A1-1}
\end{align}
where $A_{n}\lesssim B_{n}$ means that $\sup_{n}(A_{n}/B_{n})<\infty$. Let $0<\xi<\alpha/(1-\alpha)$. Note that $-\alpha/2+(1-\alpha)\xi/2<0$, and therefore, $-\delta_1+(1-\alpha)\xi/2<0$. Then, for the first term of the right-hand side in (\ref{eq:A1-1}), it follows from (C6) that
\begin{align}
&\sup_{\bm{u}\in U_{n}(r)}\left\{\frac{| \bm{u}|^{2}}{1+| \bm{u}|^{2}}
\left|\frac{1}{n}\sum_{i=1}^{n}g_{i}''(\bm{\beta}^*)+\bm{J}\right|\right\} \nonumber \\
&=n^{\delta_{1}}\left|\frac{1}{n}\sum_{i=1}^{n}g_{i}''(\bm{\beta}^*)+\bm{J}\right|
\sup_{\bm{u}\in U_{n}(r)}\left(\frac{|\bm{u}|^{2}}{1+|\bm{u}|^{2}}
\frac{|\bm{u}|^{\xi}|\bm{u}|^{-\xi}}{n^{\delta_1}}\right)
\lesssim n^{-\delta_1+(1-\alpha)\xi/2}r^{-\xi}\lesssim r^{-\xi}.
\label{eq:A1-2}
\end{align}
In addition, for the second and third terms of the right-hand side in (\ref{eq:A1-1}), we have
\begin{align}
&\sup_{\bm{u}\in U_{n}(r)}\left\{\frac{|\bm{u}|^{2}}{1+|\bm{u}|^{2}}\frac{|\bm{u}|}{n^{1/2}}\int_{0}^{1}\int_{0}^{1}
\left|\frac{1}{n}\sum_{i=1}^{n}g_{i}'''\left(\bm{\beta}^*+\frac{\bm{u}st}{n^{1/2}}\right)\right|
{\rm d}t{\rm d}s+\frac{|\bm{u}|}{1+|\bm{u}|^{2}}\right\} \nonumber \\
&\lesssim\sup_{\bm{u}\in U_{n}(r)}\left(\frac{|\bm{u}|^{2}}{1+|\bm{u}|^{2}}
\frac{|\bm{u}|^{\xi}|\bm{u}|^{-\xi}}{n^{\alpha/2}}+\frac{|\bm{u}|}{1+|\bm{u}|^{2}}\right) 
\lesssim n^{-\alpha/2+(1-\alpha)\xi/2}r^{-\xi}+r^{-1}\lesssim r^{-\xi}.
\label{eq:A1-3}
\end{align}
Letting $\gamma_{1}\in (0,\xi)$, it can be seen that (A2) holds from (\ref{eq:A1-1}), (\ref{eq:A1-2}), and (\ref{eq:A1-3}).

Finally, let us consider (A3). From Burkholder's and Jensen's inequalities, we have 
\begin{align}
\sup_{n>0}{\rm E}[|\bm{s}_n|^{N_1}]
&\leq\sup_{n>0}{\rm E}\left[\ \max_{k\leq n}\ \left|\sum_{i=1}^{k}\frac{g_i'(\bm{\beta}^*)}{n^{1/2}}\right|^{N_{1}}\right]\nonumber \\
&\lesssim \sup_{n>0}{\rm E}\left[\left\{\sum_{i=1}^{n}\frac{g_i'(\bm{\beta}^*)^2}{n}\right\}^{N_{1}/2}\right]
\leq\sup_{n>0}{\rm E}\left[\frac{1}{n}\sum_{i=1}^{n}\left|g'_{i}(\bm{\beta}^*)\right|^{N_{1}}\right]<\infty \label{eq:A3-1}
\end{align}
for $N_{1}=L(1-\gamma_{1})^{-1}\geq 2$. Let us fix $\gamma_{2}$ and $c_{2}$ such that $\alpha\chi_2<c_2<1-2\gamma_2<\min\{2\delta_2,1\}$. Since $(A+B)^{N_{2}}\lesssim A^{N_{2}}+B^{N_{2}}$ when $A$ and $B$ are positive and $N_{2}=L(1-2\gamma_{2}-c_{2})^{-1}\geq 2$, it follows from (C7) that
\begin{align}
&\sup_{n>0}{\rm E}\left[\sup_{\bm{\beta}\in \mathcal{B}}\left[n^{1/2-\gamma_{2}}\left|\frac{1}{n}\sum_{i=1}^{n}\left\{g_{i}(\bm{\beta}^*)-g_{i}(\bm{\beta})\right\}-h(\bm{\beta})\right|\right]^{N_{2}}\right]
\nonumber\\
&\lesssim\sup_{n>0}{\rm E}\left[\sup_{\bm{\beta}\in\mathcal{B}}\left\{n^{1/2-\gamma_{2}}
\left|\frac{1}{n}\sum_{i=1}^{n}\bm{y}^{\rm T}_{i}\bm{X}_{i}(\bm{\beta}^*-\bm{\beta})
-\int_{\mathcal{X}}a'(\bm{X}\bm{\beta}^*)^{\rm T}\bm{X}(\bm{\beta}^*-\bm{\beta})\mu({\rm d}\bm{X})\right|
\right\}^{N_{2}}\right] \nonumber \\
&\phantom{\lesssim}+\sup_{n>0}\sup_{\bm{\beta}\in \mathcal{B}}\left[n^{1/2-\gamma_{2}}
\left|\frac{1}{n}\sum_{i=1}^{n}\{a(\bm{X}_{i}\bm{\beta}^*)-a(\bm{X}_{i}\bm{\beta})\}
-\int_{\mathcal{X}}\{a(\bm{X}\bm{\beta}^*)-a(\bm{X}_{i}\bm{\beta})\}\mu({\rm d}\bm{X})\right|
\right]^{N_{2}}
\nonumber\\
&<\infty.
\label{eq:A3-2}
\end{align}
Further, we obtain from the precompactness of ${\cal B}$ that 
\begin{align}
\sup_{n>0}\sup_{\bm{\beta}\in \mathcal{B}}\left[n^{1/2-\gamma_{2}}\left|\frac{1}{n^{1/2}}
\sum_{j=1}^{p}\left\{p_{\lambda}(\beta_{j})-p_{\lambda}(\beta_{j}^*)\right\}\right|\right]^{N_{2}}<\infty. 
\label{eq:A3-3}
\end{align}
Hence, it can be seen that (A3) holds from (\ref{eq:A3-1}), (\ref{eq:A3-2}), and (\ref{eq:A3-3}). 

Now let us summarize the above discussion. 
\begin{thm}
Under conditions (C1)--(C7), moment convergence (\ref{eq;w_moment}) holds.
\end{thm}

By looking at the derivation of Theorem \ref{thm2} carefully, we can see that the second term in (\ref{eq;ic}) can be rewritten as
\begin{align}
{\rm E}\left[\bm{u}_{n}^{{\rm T}}\bm{s}_{n}\right]
-{\rm E}\left[\bm{u}_{n}^{{\rm T}}\{\bm{J}_{n}(\bm{\beta}^{\dagger})
-\bm{J}_{n}(\bm{\beta}^{\ddagger})\}\bm{u}_{n}\right]/2.
\label{biasorigin}
\end{align}
Let $\delta\in (0,L/2-1)$.
For the first term in (\ref{biasorigin}), it follows from the Cauchy-Schwarz inequality, (\ref{eq;w_moment}), and (\ref{eq:A3-1}) that
\begin{align*}
\sup_{n>0}{\rm E}[|\bm{u}_{n}^{{\rm T}}\bm{s}_{n}|^{1+\delta}]
\leq\left(\sup_{n>0}{\rm E}[|\bm{u}_{n}|^{2(1+\delta)}]\right)^{1/2}
\left(\sup_{n>0}{\rm E}[|\bm{s}_{n}|^{2(1+\delta)}]\right)^{1/2}<\infty.
\end{align*}
In addition, for the second term in (\ref{biasorigin}), it follows from (\ref{eq;w_moment}) that
\begin{align*}
\sup_{n>0}{\rm E}\left[|\bm{u}_{n}^{{\rm T}}\{\bm{J}_{n}(\bm{\beta}^{\dagger})
-\bm{J}_{n}(\bm{\beta}^{\ddagger})\}\bm{u}_{n}|^{1+\delta}\right]/2
\leq\chi_3\sup_{n>0}{\rm E}[|\bm{u}_{n}|^{2(1+\delta)}]<\infty,
\end{align*}
where $\chi_3$ is the supremum of the largest eigenvalue of $\bm{J}_{n}(\bm{\beta})$ over ${\cal B}$. These uniform integrabilities assure the convergence of (\ref{biasorigin}) to ${\rm E}[\tilde{\bm{u}}^{\rm T}\bm{s}]$. 

\section{Simulation study}
\label{sec;Simulation}
We conducted simulation studies to check the performance of tuning parameter selection based on the AIC in (\ref{eq;aic}). Concretely speaking, we considered a linear regression setting (Linear) and a Logistic regression setting (Logistic) and compared the performances of AIC and CV. As regularization methods, we used the Bridge ($q=0.2$), SCAD, and MCP.

We assessed the performance in terms of the second term of the Kullback-Leibler divergence:
\begin{align*}
{\rm KL}=-\tilde{{\rm E}}\left[\sum_{i=1}^{n}\tilde{g}_{i}(\hat{\bm{\beta}}_{\hat{\lambda}})\right],
\end{align*}
where $\hat{\lambda}$ is the value of the tuning parameter given by each of the criteria, and we evaluated the expectation using an empirical mean of 500 samples. We interpreted that a criterion giving a small KL value is good. Although the original aim of AIC is to minimize KL, as a secondary index for the assessment, we also determined the number of false positives and false negatives:
\begin{align*}
{\rm FP}=|\{j;\;\hat{\beta}_{j}\neq 0\wedge \beta^{*}_{j}=0\}|\;\;\;\;\;{\rm and}\;\;\;\;\;
{\rm FN}=|\{j;\;\hat{\beta}_{j}=0\wedge \beta^{*}_{j}\neq 0\}|,
\end{align*}
for each of the criteria.

The AICs we used included the one corresponding to the case $0<q<1$ in (\ref{eq;aic}) for the Bridge and the one corresponding to the case $q=1$ in (\ref{eq;aic}) for SCAD and MCP. Note that the log-likelihood function $g_{i}(\bm{\beta})$ for a linear or a logistic regression setting is expressed as
\begin{align*}
y_{i}\bm{X}_{i}\bm{\beta}-\bm{\beta}^{{\rm T}}\bm{X}_{i}^{{\rm T}}\bm{X}_{i}\bm{\beta}-y_{i}^{2}
\;\;\;\;\;{\rm or}\;\;\;\;\;
y_{i}\bm{X}_{i}\bm{\beta}-\log\{1+\exp(\bm{X}_{i}\bm{\beta})\},
\end{align*}
and $\bm{J}_{n}(\bm{\beta})$ needed for evaluating $\hat{K}$ can be expressed as
\begin{align*}
\frac{1}{n}\sum_{i=1}^{n}\bm{X}_{i}^{{\rm T}}\bm{X}_{i}
\;\;\;\;\;{\rm or}\;\;\;\;\;
\frac{1}{n}\sum_{i=1}^{n}\frac{\exp(\bm{X}_{i}\bm{\beta})}{\{1+\exp(\bm{X}_{i}\bm{\beta})\}^{2}}\bm{X}_{i}^{{\rm T}}\bm{X}_{i}.
\end{align*}

The simulation settings were as follows. As $p$-dimensional regressors $\bm{X}_{i},\;(i=1,2,\ldots,n)$, we used vectors obtained from the multivariate Gaussian distribution ${\rm N}(\bm{0},\bm{\Sigma})$, where $\bm{\Sigma}$ is $(p\times p)$-covariance matrix whose $(i,j)$-th element was set to $0.5^{|i-j|}$. The true coefficient vector $\bm{\beta}^{*}$ was
\begin{align*}
\bm{\beta}^{*}
=(\beta_{1}^{*}\bm{1}_{k}^{{\rm T}},\beta_{2}^{*}\bm{1}_{k}^{{\rm T}},\bm{0}_{p-2k}^{{\rm T}})^{{\rm T}},
\end{align*}
where $\bm{1}_{k}$ and $\bm{0}_{p-2k}$ respectively denote a $k$-dimensional one-vector and a $(p-2k)$-dimensional zero-vector. In addition, $(\beta_{1}^{*},\beta_{2}^{*})$ was set to (0.1,0.5) or (0.2,1) in the linear regression setting and (0.5,1.5) or (1,2) in the logistic regression setting, and seven cases of the three-tuple $(p,k,n)$ were considered: (8,2,50), (8,2,100), (8,2,150), (8,1,100), (8,3,100), (12,3,100), and (16,4,100). We used the local quadratic approximation in \citet{FanLi01} for the parameter estimation and conducted fifty simulations.

Tables \ref{tb;Bridge}, \ref{tb;SCAD}, and \ref{tb;MCP} show the results for the Bridge, SCAD, and MCP, respectively. Each table lists the averages and standard deviations of KL, as well as the averages of FP and FN, for the linear and the logistic regression settings. Let us look at the main index in Table \ref{tb;Bridge}. While CV gives a smaller KL value than AIC does in about half the cases, the differences between the two values are small. On the other hand, in the cases in which AIC gives a smaller KL value than CV does, the differences tend to be large. Next, let us look at the sub indices FP and FN. In the logistic setting, the FP values are almost 0 while those of FN are rather large. That is, we can say that CV causes an imbalance. So long as there is no special reason of give importance on the FP, it will be natural to use the AIC. In Tables \ref{tb;SCAD} and \ref{tb;MCP}, AIC and CV give almost the same values of KL in the linear setting. On the other hand, in the logistic setting, AIC is clearly superior to CV in many cases. On the whole, we can conclude that the AIC in (\ref{eq;aic}) is better than CV.

\begin{table}[htbp]
\begin{center}
\begin{tabular}{lll|cccccc}
\hline
&&&\multicolumn{3}{c}{Case 1}&\multicolumn{3}{c}{Case 2} \\
Model&$(p,k,n)$&&KL\;(sd)&FP&FN&KL\;(sd)&FP&FN \\ \hline
Linear&(8,2,50)&CV&0.676\;(0.019)&0.30&1.58&0.645\;(0.026)&0.30&1.29 \\
&&AIC& 0.679\;(0.018)&0.09&1.77&0.649\;(0.022)&0.11&1.55 \\ \cline{2-9}
&(8,2,100)&CV& 0.670\;(0.016)&0.31&1.31&0.631\;(0.018)&0.28&1.05 \\
&&AIC& 0.672\;(0.015)&0.05&1.61&0.634\;(0.018)&0.07&1.27 \\ \cline{2-9}
&(8,2,150)&CV&0.666\;(0.014)&0.32&1.24&0.632\;(0.012)&0.40&0.86 \\
&&AIC&0.666\;(0.013)&0.10&1.45&0.636\;(0.014)&0.04&1.17 \\ \cline{2-9}
&(8,1,100)&CV&0.687\;(0.008)&0.46&0.75&0.658\;(0.017)&0.75&0.45  \\
&&AIC&0.687\;(0.009)&0.12&0.81&0.658\;(0.016)&0.13&0.54  \\ \cline{2-9}
&(8,3,100)&CV&0.655\;(0.014)&0.24&1.86&0.615\;(0.020)&0.24&1.40  \\
&&AIC&0.659\;(0.012)&0.03&2.34&0.626\;(0.019)&0.04&2.19  \\ \cline{2-9}
&(12,3,100)&CV& 0.662\;(0.014)&0.47&1.91&0.617\;(0.021)&0.46&1.64 \\
&&AIC& 0.665\;(0.014)&0.15&2.38&0.624\;(0.018)&0.06&2.17 \\ \cline{2-9}
&(16,4,100)&CV& 0.652\;(0.021)&0.41&3.03&0.610\;(0.024)&0.69&2.47 \\
&&AIC& 0.652\;(0.017)&0.12&3.28&0.618\;(0.021)&0.12&2.98 \\ \hline
Logistic&(8,2,50)&CV& 0.462\;(0.061)& 0.01& 1.28& 0.406\;(0.070)& 0.04& 1.21 \\
&&AIC& 0.473\;(0.153)& 0.33& 0.69& 0.417\;(0.129)& 0.40& 0.40\\ \cline{2-9}
&(8,2,100)&CV& 0.419\;(0.044)& 0.01& 1.04& 0.348\;(0.047)& 0.00& 0.92 \\
&&AIC& 0.398\;(0.050)& 0.31& 0.43& 0.307\;(0.035)& 0.50& 0.19 \\ \cline{2-9}
&(8,2,150)&CV& 0.394\;(0.024)& 0.00& 0.94& 0.307\;(0.033)& 0.01& 0.67 \\
&&AIC& 0.376\;(0.018)& 0.43& 0.33& 0.271\;(0.018)& 0.41& 0.11 \\ \cline{2-9}
&(8,1,100)&CV& 0.495\;(0.029)& 0.00& 0.42& 0.411\;(0.021)& 0.00& 0.22 \\
&&AIC& 0.513\;(0.033)& 0.61& 0.21& 0.423\;(0.035)& 0.63& 0.02 \\ \cline{2-9}
&(8,3,100)&CV& 0.408\;(0.047)& 0.00& 1.92& 0.348\;(0.053)& 0.00& 1.74 \\
&&AIC& 0.346\;(0.042)& 0.22& 0.78& 0.272\;(0.087)& 0.35& 0.32 \\ \cline{2-9}
&(12,3,100)&CV& 0.384\;(0.031)& 0.01& 1.82& 0.376\;(0.056)& 0.00& 1.68 \\
&&AIC& 0.397\;(0.134)& 0.75& 0.58& 0.346\;(0.112)& 0.73& 0.35 \\ \cline{2-9}
&(16,4,100)&CV& 0.392\;(0.048)& 0.01& 2.72& 0.407\;(0.045)& 0.00& 2.66 \\
&&AIC& 0.414\;(0.122)& 1.19& 1.05& 0.379\;(0.137)& 1.17& 0.60 \\ \hline
\end{tabular}
\caption{Comparison of CV and AIC in (\ref{eq;aic}) for the Bridge penalty. The true parameter vector $(\beta_{1}^{*},\beta_{2}^{*})$ is (0.1,0.5) for Case 1 and (0.2,1) for Case 2 in the linear regression setting and (0.5,1.5) and (1,2) in the logistic regression setting.}
\label{tb;Bridge}
\end{center}
\end{table}

\begin{table}[htbp]
\begin{center}
\begin{tabular}{lll|cccccc}
\hline
&&&\multicolumn{3}{c}{Case 1}&\multicolumn{3}{c}{Case 2} \\
Model&$(p,k,n)$&&KL\;(sd)&FP&FN&KL\;(sd)&FP&FN \\ \hline
Linear&(8,2,50)&CV& 0.557\;(0.050)&0.69&0.49&0.563\;(0.039)&0.87&0.20 \\
&&AIC& 0.566\;(0.055)&0.60&0.59&0.582\;(0.056)&0.95&0.20 \\ \cline{2-9}
&(8,2,100)&CV& 0.521\;(0.020)&1.01&0.27&0.518\;(0.031)&0.93&0.11 \\
&&AIC& 0.524\;(0.025)&0.92&0.28&0.519\;(0.028)&0.91&0.15 \\ \cline{2-9}
&(8,2,150)&CV& 0.531\;(0.013)&0.76&0.24&0.567\;(0.012)&1.05&0.03\\
&&AIC& 0.534\;(0.015)&0.70&0.26&0.569\;(0.013)&0.89&0.03 \\ \cline{2-9}
&(8,1,100)&CV& 0.526\;(0.021)&1.24&0.19&0.500\;(0.020)&1.26&0.06 \\
&&AIC& 0.526\;(0.025)&1.05&0.24&0.503\;(0.023)&1.13&0.06 \\ \cline{2-9}
&(8,3,100)&CV& 0.491\;(0.020)&0.49&0.41&0.555\;(0.025)&0.59&0.17 \\
&&AIC& 0.492\;(0.021)&0.43&0.51&0.555\;(0.027)&0.48&0.22 \\ \cline{2-9}
&(12,3,100)&CV& 0.504\;(0.020)&1.16&0.37&0.556\;(0.023)&1.33&0.15 \\
&&AIC& 0.509\;(0.028)&1.15&0.38&0.561\;(0.026)&1.23&0.16 \\ \cline{2-9}
&(16,4,100)&CV& 0.550\;(0.030)&1.54&0.66&0.565\;(0.029)&1.80&0.15 \\
&&AIC& 0.557\;(0.035)&1.39&0.66&0.573\;(0.031)&1.44&0.24 \\ \hline
Logistic&(8,2,50)&CV& 0.506\;(0.032)&0.04&0.82&0.493\;(0.023)&0.06&0.59 \\
&&AIC& 0.477\;(0.117)&0.76&0.56&0.511\;(0.184)&0.48&0.54 \\ \cline{2-9}
&(8,2,100)&CV& 0.476\;(0.017)&0.07&0.69&0.426\;(0.018)&0.04&0.20 \\
&&AIC& 0.446\;(0.059)&0.78&0.41&0.321\;(0.037)&0.52&0.25 \\ \cline{2-9}
&(8,2,150)&CV& 0.451\;(0.015)&0.05&0.41&0.394\;(0.015)&0.06&0.13\\
&&AIC& 0.411\;(0.021)&1.09&0.18&0.301\;(0.025)&0.95&0.08 \\ \cline{2-9}
&(8,1,100)&CV& 0.541\;(0.017)&0.15&0.14&0.454\;(0.024)&0.07&0.06 \\
&&AIC& 0.542\;(0.036)&1.40&0.09&0.406\;(0.029)&1.01&0.04 \\ \cline{2-9}
&(8,3,100)&CV& 0.431\;(0.017)&0.05&1.09&0.423\;(0.015)&0.05&0.54 \\
&&AIC& 0.339\;(0.043)&0.38&0.66&0.314\;(0.056)&0.19&0.55 \\ \cline{2-9}
&(12,3,100)&CV& 0.449\;(0.014)&0.03&0.95&0.420\;(0.015)&0.03&0.53 \\
&&AIC& 0.413\;(0.093)&1.44&0.46&0.349\;(0.086)&0.86&0.59 \\ \cline{2-9}
&(16,4,100)&CV& 0.436\;(0.013)&0.08&1.50&0.423\;(0.018)&0.06&1.19 \\
&&AIC& 0.438\;(0.115)&1.52&0.99&0.356\;(0.080)&0.87&1.11 \\ \hline
\end{tabular}
\caption{Comparison of CV and AIC in (\ref{eq;aic}) for the SCAD penalty. The true parameter vector $(\beta_{1}^{*},\beta_{2}^{*})$ is (0.1,0.5) for Case 1  and (0.2,1) for Case 2 in the linear regression setting and (0.5,1.5) and (1,2) in the logistic regression setting.}
\label{tb;SCAD}
\end{center}
\end{table}

\begin{table}[htbp]
\begin{center}
\begin{tabular}{lll|cccccc}
\hline
&&&\multicolumn{3}{c}{Case 1}&\multicolumn{3}{c}{Case 2} \\
Model&$(p,k,n)$&&KL\;(sd)&FP&FN&KL\;(sd)&FP&FN \\ \hline
Linear&(8,2,50)&CV& 0.545\;(0.047)&0.82&0.42&0.556\;(0.046)&0.79&0.23 \\
&&AIC& 0.545\;(0.047)&0.67&0.49&0.557\;(0.046)&0.71&0.29 \\ \cline{2-9}
&(8,2,100)&CV& 0.558\;(0.020)&0.79&0.38&0.527\;(0.023)&0.86&0.13 \\
&&AIC& 0.560\;(0.026)&0.64&0.39&0.530\;(0.027)&0.92&0.13 \\ \cline{2-9}
&(8,2,150)&CV& 0.520\;(0.017)&0.91&0.31&0.518\;(0.015)&0.94&0.10\\
&&AIC& 0.521\;(0.018)&0.71&0.38&0.519\;(0.015)&0.84&0.11 \\ \cline{2-9}
&(8,1,100)&CV& 0.502\;(0.015)&1.02&0.25&0.539\;(0.023)&1.03&0.15 \\
&&AIC& 0.503\;(0.018)&0.88&0.27&0.540\;(0.024)&0.99&0.14 \\ \cline{2-9}
&(8,3,100)&CV& 0.553\;(0.021)&0.33&0.53&0.508\;(0.028)&0.62&0.10 \\
&&AIC& 0.556\;(0.023)&0.30&0.61&0.510\;(0.029)&0.49&0.16 \\ \cline{2-9}
&(12,3,100)&CV& 0.523\;(0.023)&1.24&0.57&0.578\;(0.030)&1.45&0.17 \\
&&AIC& 0.525\;(0.024)&1.02&0.57&0.582\;(0.028)&1.39&0.19 \\ \cline{2-9}
&(16,4,100)&CV& 0.530\;(0.029)&1.72&0.72&0.563\;(0.035)&1.73&0.28 \\
&&AIC& 0.532\;(0.031)&1.45&0.72&0.565\;(0.036)&1.53&0.34 \\ \hline
Logistic&(8,2,50)&CV& 0.493\;(0.037)&0.04&1.04&0.453\;(0.035)&0.06&0.81 \\
&&AIC& 0.514\;(0.159)&0.59&0.59&0.383\;(0.090)&0.41&0.52 \\ \cline{2-9}
&(8,2,100)&CV& 0.447\;(0.023)&0.02&0.65&0.397\;(0.025)&0.02&0.47 \\
&&AIC& 0.418\;(0.043)&0.79&0.29&0.323\;(0.029)&0.54&0.21 \\ \cline{2-9}
&(8,2,150)&CV& 0.423\;(0.017)&0.04&0.54&0.367\;(0.019)&0.01&0.17\\
&&AIC& 0.390\;(0.019)&0.88&0.21&0.308\;(0.020)&0.94&0.09 \\ \cline{2-9}
&(8,1,100)&CV& 0.529\;(0.020)&0.10&0.23&0.448\;(0.021)&0.13&0.08 \\
&&AIC& 0.530\;(0.036)&0.83&0.17&0.429\;(0.027)&1.06&0.06 \\ \cline{2-9}
&(8,3,100)&CV& 0.429\;(0.020)&0.01&1.09&0.409\;(0.031)&0.02&0.97 \\
&&AIC& 0.362\;(0.056)&0.33&0.70&0.312\;(0.075)&0.16&0.73 \\ \cline{2-9}
&(12,3,100)&CV& 0.423\;(0.027)&0.01&1.10&0.401\;(0.017)&0.01&0.99 \\
&&AIC& 0.389\;(0.070)&1.02&0.66&0.352\;(0.075)&0.91&0.65 \\ \cline{2-9}
&(16,4,100)&CV& 0.426\;(0.022)&0.02&1.92&0.411\;(0.017)&0.02&1.54 \\
&&AIC& 0.440\;(0.136)&1.79&0.94&0.345\;(0.107)&1.31&1.02 \\ \hline
\end{tabular}
\caption{Comparison of CV and AIC in (\ref{eq;aic}) for the MCP penalty. The true parameter vector $(\beta_{1}^{*},\beta_{2}^{*})$ is (0.1,0.5) for Case 1 and (0.2,1) for Case 2 in the linear regression setting and (0.5,1.5) and (1,2) in the logistic regression setting.}
\label{tb;MCP}
\end{center}
\end{table}

\section{Discussion}
\label{sec;Discussion}
Although \citet{NinKaw14} derived an information criterion for the Lasso in generalized linear models on the basis of the original definition of AIC, which is an asymptotically unbiased estimator of the Kullback-Leibler divergence, they used an asymptotic setting wherein estimation consistency is not assured. In addition, the Lasso itself has a problem in that efficiency is not necessarily high because it shrinks the estimator to the zero vector too much. As a way of dealing with these problems, we derived an information criterion for non-concave penalized maximum likelihood methods including the Bridge, SCAD, and MCP, which are known to be more efficient than the Lasso, on the basis of the original definition of AIC in a setting in which estimation consistency is assured. The AIC in (\ref{eq;aic}) is the only criterion for such non-concave penalized maximum likelihood methods that has the same roots as those of the classic information criteria. Its bias term, including its coefficient, is determined. Therefore, unlike the information criteria that assure model selection consistency, it allows us to perform a model selection without any arbitrariness.

It has been shown through simulation studies that the performance of the AIC in (\ref{eq;aic}) is almost the same as or better than that of the CV. In terms of computational cost, AIC is clearly better than CV in the Bridge-type regularization method because of its simple expression. This fact is a significant advantage when handling large-scale data.

Although the number of tuning parameters to be selected is only one, we can extend our result to regularization methods that have several tuning parameters, such as SELO (\citealt{DicHuaLin12}). In addition, although we used the natural link function for our generalized linear models, it is possible to treat different link functions given certain regularity conditions. In this study, we derived the AIC based on statistical asymptotic theory for which the dimension of the parameter vector is fixed and the sample size diverges. On the other hand, it is becoming important to analyze high-dimensional data wherein the dimension of the parameter vector is comparable to the sample size. Also for such high-dimensional data, we expect that the AIC-type information criterion will work well from the viewpoint of efficiency. In fact, \citet{ZhaLiTsa10} has shown that, when the dimension of the parameter vector increases with the sample size, their criterion close to the proposed information criterion has an asymptotic loss efficiency in a sparse setting under certain conditions. It will be important in terms of both theory and practice to show that the proposed information criterion has a similar asymptotic property.

\begin{appendix}
\section{Proofs}
\subsection{Proof of Lemma \ref{lem2}}
\label{app;lem2}
From (R1), the first term in the right-hand side of (\ref{defmu}) converges in probability to $h(\bm{\beta})$ for each $\bm{\beta}$. In addition, from the convexity of $\mu_{n}(\bm{\beta})$ with respect to $\bm{\beta}$, we have
\begin{align*}
\sup_{\bm{\beta}\in K}\left|\frac{1}{n}\sum_{i=1}^{n}\{g_{i}(\bm{\beta}^{*})-g_{i}(\bm{\beta})\}-h(\bm{\beta})\right|\stackrel{{\rm p}}{\to} 0
\end{align*}
for any compact set $K$ (\citealt{AndGil82,Pol91}). Accordingly, we have
\begin{align}
\sup_{\bm{\beta}\in K}|\mu_{n}(\bm{\beta})-h(\bm{\beta})|\stackrel{{\rm p}}{\to}0.
\label{eq;unif}
\end{align}
Note that in the following inequality,
\begin{align*}
\mu_{n}(\bm{\beta})\geq\frac{1}{n}\sum_{i=1}^{n}\{g_{i}(\bm{\beta}^{*})-g_{i}(\bm{\beta})\} \equiv \mu_{n}^{(0)}(\bm{\beta}),
\end{align*}
the argmin of the right-hand side is the maximum likelihood estimator and is ${\rm O}_{{\rm p}}(1)$. Also note that for some $M\ (>0)$,
\begin{align*}
{\rm P}(|\hat{\bm{\beta}}_{\lambda}|>M)
\leq {\rm P}\left(\inf_{|\bm{\beta}|>M}\mu_{n}(\bm{\beta})\leq \mu_{n}(\bm{0})\right)
\leq {\rm P}\left(\inf_{|\bm{\beta}|>M}\mu_{n}^{(0)}(\bm{\beta})\leq \mu_{n}^{(0)}(\bm{0})\right)
\end{align*}
because $p_{\lambda}(0)=0$ from (C4). Therefore, we have
\begin{align}
\hat{\bm{\beta}}_{\lambda}=\underset{\bm{\beta}\in{\cal B}}{{\rm agmin}}\;\mu_{n}(\bm{\beta})={\rm O}_{{\rm p}}(1).
\label{eq;Op}
\end{align}
From (\ref{eq;unif}) and (\ref{eq;Op}), we obtain
\begin{align*}
\hat{\bm{\beta}}_{\lambda}
=\underset{\bm{\beta}\in{\cal B}}{{\rm argmin}}\;\mu_{n}(\bm{\beta})
\stackrel{{\rm p}}{\to}\underset{\bm{\beta}\in{\cal B}}{{\rm argmin}}\;h(\bm{\beta})
=\bm{\beta}^{*}.
\end{align*}

\subsection{Proof of (\ref{prelem3_2})}
\label{app;lem3}
Let $\bm{u}=\tilde{\bm{u}}_{n}+l\bm{w}$, where $\bm{w}$ is a unit vector, and let $l\in(\delta,\xi)$. The strong convexity of $\eta_{n}(\bm{u})$ implies
\begin{align*}
(1-\delta/l)\eta_{n}(\tilde{\bm{u}}_{n})+(\delta/l)\eta_{n}(\bm{u})
> \eta_{n}(\tilde{\bm{u}}_{n}+\delta\bm{w}),
\end{align*}
and we thus have
\begin{align*}
(\delta/l)\{\nu_{n}(\bm{u})-\nu_{n}(\tilde{\bm{u}}_{n})\}
> & \nu_{n}(\tilde{\bm{u}}_{n}+\delta\bm{w})-\nu_{n}(\tilde{\bm{u}}_{n}) \\
& +(1-\delta/l)\phi_{n}(\tilde{\bm{u}}_{n})+(\delta/l)\phi_{n}(\bm{u})-\phi_{n}(\tilde{\bm{u}}_{n}+\delta\bm{w}) \\
&+(1-\delta/l)\psi_{n}(\tilde{\bm{u}}_{n}^{\dagger})+(\delta/l)\psi_{n}(\bm{u}^{\dagger})-\psi_{n}(\tilde{\bm{u}}_{n}^{\dagger}+\delta\bm{w}^{\dagger}).
\end{align*}
Since it follows that
\begin{align*}
&\nu_{n}(\tilde{\bm{u}}_{n}+\delta\bm{w})-\nu_{n}(\tilde{\bm{u}}_{n}) \\
&=\{\nu_{n}(\tilde{\bm{u}}_{n}+\delta\bm{w})-\tilde{\nu}_{n}(\tilde{\bm{u}}_{n}+\delta\bm{w})\}+\{\tilde{\nu}_{n}(\tilde{\bm{u}}_{n}+\delta\bm{w})-\tilde{\nu}_{n}(\tilde{\bm{u}}_{n})\}+\{\tilde{\nu}_{n}(\tilde{\bm{u}}_{n})-\nu_{n}(\tilde{\bm{u}}_{n})\} \\
&\geq\Upsilon_{n}(\delta)-2\Delta_{n}(\delta),
\end{align*}
we obtain from (\ref{prelem3_1_phi}) and (\ref{prelem3_1_psi}) that, for any $\varepsilon\;(>0)$, 
\begin{align*}
(\delta/l)\{\nu_{n}(\bm{u})-\nu_{n}(\tilde{\bm{u}}_{n})\}
> \Upsilon_{n}(\delta)-2\Delta_{n}(\delta)-\varepsilon
\end{align*}
for sufficiently large $n$ and sufficiently small $\gamma$. If $2\Delta_{n}(\delta)+\varepsilon<\Upsilon_{n}(\delta)$, then $\nu_{n}(\bm{u})\geq \nu_{n}(\tilde{\bm{u}}_{n})$ for any $\bm{u}$ such that $|\bm{u}^{\dagger}|\leq \gamma$ and $\delta\leq|\bm{u}-\tilde{\bm{u}}_{n}|\leq\xi$. This means $\bm{u}_{n}$ must satisfy $|\bm{u}_{n}^{\dagger}|> \gamma$ or $|\bm{u}_{n}-\tilde{\bm{u}}_{n}|\not\in [\delta,\xi]$ in order for $\bm{u}_{n}$ to be the argmin of $\nu_{n}(\bm{u})$. Hence, we obtain (\ref{prelem3_2}).

\subsection{Proof of (\ref{eq;Rad})}
\label{app;Rad}
Let us consider a random function $\mu_{n}(\bm{\beta})$ in (\ref{defmu}). Since $p_{\lambda}(0)=0$ from (C4), we have
\begin{align*}
\mu_{n}(\hat{\bm{\beta}}_{\lambda})
=&-n^{-1/2}\bm{s}_{n}^{{\rm T}}(\hat{\bm{\beta}}_{\lambda}-\bm{\beta}^{*})+(\hat{\bm{\beta}}_{\lambda}-\bm{\beta}^{*})^{{\rm T}}\bm{J}_{n}(\tilde{\bm{\beta}})(\hat{\bm{\beta}}_{\lambda}-\bm{\beta}^{*})/2 \\
&+n^{-1/2}\sum_{j\in{\cal J}^{(1)}}p_{\lambda}(\hat{\beta}_{\lambda,j})+n^{-1/2}\sum_{j\in{\cal J}^{(2)}}p'_{\lambda}(\beta_{j}^{*})(\hat{\beta}_{\lambda,j}-\beta_{j}^{*})\{1+{\rm o}_{{\rm p}}(1)\},
\end{align*}
where $\tilde{\bm{\beta}}$ is a vector on the segment from $\hat{\bm{\beta}}_{\lambda}$ to $\bm{\beta}^{*}$. Then, we have
\begin{align*}
0
\geq \mu_{n}(\hat{\bm{\beta}}_{\lambda})-\mu_{n}(\bm{\beta}^{*})
\geq {\rm O}_{{\rm p}}(n^{-1/2}|\hat{\bm{\beta}}_{\lambda}-\bm{\beta}^{*}|)+(\hat{\bm{\beta}}_{\lambda}-\bm{\beta}^{*})^{{\rm T}}\bm{J}_{n}(\tilde{\bm{\beta}})(\hat{\bm{\beta}}_{\lambda}-\bm{\beta}^{*})/2
\end{align*}
because $\bm{s}_{n}={\rm O}_{{\rm p}}(1)$. From (C2), $\bm{J}_{n}(\tilde{\bm{\beta}})$ is positive definite for sufficiently large $n$, and therefore, it follows that
\begin{align}
\hat{\bm{\beta}}_{\lambda}-\bm{\beta}^{*}={\rm O}_{{\rm p}}(n^{-1/2}).
\label{eq;Op(1)}
\end{align}
Let us express $\mu_{n}(\bm{\beta})$ by $\mu_{n}(\bm{\beta}^{(1)},\bm{\beta}^{(2)})$. Because $0\geq\mu_{n}(\hat{\bm{\beta}}_{\lambda}^{(1)},\hat{\bm{\beta}}_{\lambda}^{(2)})-\mu_{n}(\bm{0},\hat{\bm{\beta}}_{\lambda}^{(2)})$, we see that
\begin{align*}
\;-n^{-1/2}\bm{s}_{n}^{(1){\rm T}}\hat{\bm{\beta}}_{\lambda}^{(1)}
+\hat{\bm{\beta}}_{\lambda}^{(1){\rm T}}\bm{J}^{(11)}_{n}(\tilde{\bm{\beta}})\hat{\bm{\beta}}_{\lambda}^{(1)}/2
+\hat{\bm{\beta}}_{\lambda}^{(1){\rm T}}\bm{J}^{(11)}_{n}(\tilde{\bm{\beta}})(\hat{\bm{\beta}}_{\lambda}^{(2)}-\bm{\beta}^{*(2)})
+n^{-1/2}\sum_{j\in{\cal J}^{(1)}}p_{\lambda}(\hat{\beta}_{\lambda,j})
\end{align*}
is non-positive. Here, we use the fact that $\sum_{j\in{\cal J}^{(1)}}p_{\lambda}(\hat{\beta}_{\lambda,j})$ reduces to $\lambda\|\hat{\bm{\beta}}_{\lambda}^{(1)}\|_{q}^{q}\{1+{\rm o}_{{\rm p}}(1)\}$ from (C4) and (\ref{eq;Op(1)}) and that $\bm{J}_{n}(\tilde{\bm{\beta}})$ is positive definite for sufficiently large $n$. Accordingly, we have
\begin{align*}
|\hat{\bm{\beta}}_{\lambda}^{(1)}|^{2}+n^{-1/2}\|\hat{\bm{\beta}}_{\lambda}^{(1)}\|_{q}^{q}\{1+{\rm o}_{{\rm p}}(1)\}\leq {\rm O}_{{\rm p}}(n^{-1/2}|\hat{\bm{\beta}}_{\lambda}^{(1)}|)
\end{align*}
and thus $\|\hat{\bm{\beta}}_{\lambda}^{(1)}\|_{q}^{q}\leq {\rm O}_{{\rm p}}(|\hat{\bm{\beta}}_{\lambda}^{(1)}|)$. Hence, we have
\begin{align}
{\rm P}(\hat{\bm{\beta}}_{\lambda}^{(1)}=\bm{0})\to1
\label{eq;zero}
\end{align}
because $0<q<1$ and $\hat{\bm{\beta}}_{\lambda}^{(1)}={\rm o}_{{\rm p}}(1)$. This implies the former in (\ref{eq;Rad}). Since $\tilde{\bm{u}}_{n}^{(2)}$ is trivially ${\rm O}_{{\rm p}}(1)$, we obtain the latter of (\ref{eq;Rad}) from (\ref{eq;Op(1)}) and (\ref{eq;zero}).

\subsection{Proof of (\ref{eq;a1-1}) and (\ref{eq;a1-2})}
\label{app;thm1}
Let $\eta_{n}(\bm{u}^{(1)},\bm{u}^{(2)})$ be the one with $q=1$ in (\ref{eq;eta_n}), and let $\tilde{\eta}_{n}(\bm{u}^{(1)},\bm{u}^{(2)})=-\bm{u}^{{\rm T}}\bm{s}_{n}+\bm{u}^{{\rm T}}\bm{J}\bm{u}/2$ in place of (\ref{eq;approx}). Then, we can obtain $\eta_{n}(\bm{u}^{(1)},\bm{u}^{(2)})=\tilde{\eta}_{n}(\bm{u}^{(1)},\bm{u}^{(2)})+{\rm o}_{{\rm p}}(1)$ by taking a Taylor expansion around $(\bm{u}^{(1)},\bm{u}^{(2)})=(\bm{0},\bm{0})$. In addition, let $\phi_{n}(\bm{u})$ and $\phi(\bm{u})$ be $\phi_{n}(\bm{u})+\psi_{n}(\bm{u}^{\dagger})$ and $\phi(\bm{u})+\psi(\bm{u}^{\dagger})$ with $q=1$ in (\ref{eq;phi_n}), (\ref{eq;psi_n}) and (\ref{eq;phi}), let $\bm{u}^{\dagger}$ be empty vector and $\psi_{n}(\bm{u}^{\dagger})=\psi(\bm{u}^{\dagger})=0$, and define $\nu_{n}(\bm{u}^{(1)},\bm{u}^{(2)})=\eta_{n}(\bm{u}^{(1)},\bm{u}^{(2)})+\phi_{n}(\bm{u})+\psi_{n}(\bm{u}^{\dagger})$ and $\tilde{\nu}_{n}(\bm{u}^{(1)},\bm{u}^{(2)})=\tilde{\eta}_{n}(\bm{u}^{(1)},\bm{u}^{(2)})+\phi(\bm{u})+\psi(\bm{u}^{\dagger})$ again. Here, note that
\begin{align*}
(\bm{u}_{n}^{(1)},\bm{u}_{n}^{(2)})
=\underset{(\bm{u}^{(1)},\bm{u}^{(2)})}{{\rm argmin}}\nu_{n}(\bm{u}^{(1)},\bm{u}^{(2)})
=(n^{1/2}\hat{\bm{\beta}}^{(1)}_{\lambda},n^{1/2}(\hat{\bm{\beta}}^{(2)}_{\lambda}-\bm{\beta}^{*(2)})).
\end{align*}
Next, because
\begin{align*}
\tilde{\nu}_{n}(\bm{u}^{(1)},\bm{u}^{(2)})
=&\|\bm{u}^{(2)}-\bm{J}^{(22)-1}\{-\bm{J}^{(21)}\bm{u}^{(1)}
+(\bm{s}_{n}^{(2)}-\bm{p}'^{(2)}_{\lambda})\}\|_{\bm{J}^{(22)}}^{2}/2 \\
&+\bm{u}^{(1){\rm T}}\bm{J}^{(1|2)}\bm{u}^{(1)}/2
-\bm{u}^{(1){\rm T}}\bm{\tau}_{\lambda}(\bm{s}_{n})
+\lambda\|\bm{u}^{(1)}\|_{1}
-\|\bm{s}_{n}^{(2)}-\bm{p}'^{(2)}_{\lambda}\|_{\bm{J}^{(22)-1}}^{2}/2,
\end{align*}
we see by using $\hat{\bm{u}}_{n}^{(1)}$ in (\ref{eq;u1}) that
\begin{align*}
(\tilde{\bm{u}}_{n}^{(1)},\tilde{\bm{u}}_{n}^{(2)})
=\underset{(\bm{u}^{(1)},\bm{u}^{(2)})}{{\rm argmin}}\tilde{\nu}_{n}(\bm{u}^{(1)},\bm{u}^{(2)})
=(\hat{\bm{u}}_{n}^{(1)},-\bm{J}^{(22)-1}\bm{J}^{(21)}\hat{\bm{u}}_{n}^{(1)}+\bm{J}^{(22)-1}(\bm{s}_{n}^{(2)}-\bm{p}'^{(2)}_{\lambda})),
\end{align*}
where we have denoted $\bm{x}^{{\rm T}}A\bm{x}$ by $\|\bm{x}\|_{A}^{2}$ for an appropriate size of matrix $A$ and vector $\bm{x}$. Now we apply Lemma \ref{lem3} and evaluate the right-hand side in (\ref{eq;lem3_ineq}). In the same way as in (\ref{eq;Delta_conv}), it follows that $\Delta_{n}(\delta)$ converges in probability to $0$. Next, the definition of $\tilde{\bm{u}}_{n}^{(1)}$ ensures that
\begin{align*}
\bm{J}^{(1|2)}\tilde{\bm{u}}_{n}^{(1)}-\bm{\tau}_{\lambda}(\bm{s}_{n})+\lambda\bm{\gamma}=\bm{0},
\end{align*}
where $\bm{\gamma}$ is a $|{\cal J}^{(1)}|$-dimensional vector such that $\gamma_{j}=1$ when $\hat{u}_{n,j}^{(1)}>0$, $\gamma_{j}=-1$ when $\hat{u}_{n,j}^{(1)}<0$, and $\gamma_{j}\in[-1,1]$ when $\hat{u}_{n,j}^{(1)}=0$. Thus, noting that $\tilde{\bm{u}}_{n}^{(1){\rm T}}\bm{\gamma}=\|\tilde{\bm{u}}_{n}^{(1)}\|_{1}$, we can write $\tilde{\nu}_{n}(\bm{u}^{(1)},\bm{u}^{(2)})-\tilde{\nu}_{n}(\tilde{\bm{u}}_{n}^{(1)},\tilde{\bm{u}}_{n}^{(2)})$ as
\begin{align}
&\|\bm{u}^{(1)}-\tilde{\bm{u}}_{n}^{(1)}\|_{\bm{J}^{(1|2)}}^{2}/2
+\lambda\sum_{j\in{\cal J}^{(1)}}\left(|u_{j}|-\gamma_{j}u_{j}\right) \nonumber \\
&+\|\bm{u}^{(2)}-\bm{J}^{(22)-1}\{-\bm{J}^{(21)}\bm{u}^{(1)}
+(\bm{s}_{n}^{(2)}-\bm{p}'^{(2)}_{\lambda})\}\|_{\bm{J}^{(22)}}^{2}/2
\label{eq:Upsilon_n}
\end{align}
after a simple calculation. Let $\bm{w}_{1}$ and $\bm{w}_{2}$ be unit vectors such that $\bm{u}^{(1)}=\tilde{\bm{u}}_{n}^{(1)}+\zeta\bm{w}_{1}$ and $\bm{u}^{(2)}=\tilde{\bm{u}}_{n}^{(2)}+(\delta^{2}-\zeta^{2})^{1/2}\bm{w}_{2}$, where $0\leq \zeta\leq \delta$. Then, letting $\rho^{(22)}$ and $\rho^{(1|2)}\;(>0)$ be half the smallest eigenvalues of $\bm{J}^{(22)}$ and $\bm{J}^{(1|2)}$, respectively, it follows that
\begin{align*}
\Upsilon_{n}(\delta)
\geq \min_{0\leq \zeta\leq \delta}\left\{ \rho^{(1|2)}\zeta^{2}+\rho^{(22)}|(\delta^{2}-\zeta^{2})^{1/2}\bm{w}_{2}+\zeta\bm{J}^{(22)-1}\bm{J}^{(21)}\bm{w}_{1}|^{2}\right\}
>0
\end{align*}
because the second term in (\ref{eq:Upsilon_n}) is non-negative. Hence, the first term on the right-hand side in (\ref{eq;lem3_ineq}) converges to 0. In addition, because $(\bm{u}_{n}^{(1)},\bm{u}_{n}^{(2)})$ is ${\rm O}_{{\rm p}}(1)$ from (\ref{eq;Op(1)}) and $(\tilde{\bm{u}}_{n}^{(1)},\tilde{\bm{u}}_{n}^{(2)})$ is also ${\rm O}_{{\rm p}}(1)$, the second term on the right-hand side in (\ref{eq:lem3}) can be made arbitrarily small by considering a sufficiently large $\xi$. Thus, we have $|\bm{u}-\tilde{\bm{u}}_{n}|={\rm o}_{{\rm p}}(1)$, and as a consequence, we obtain (\ref{eq;a1-1}) and (\ref{eq;a1-2}).

\subsection{Proof of (\ref{eq;a1-3})}
\label{app;thm2}
Because $n^{1/2}\hat{\bm{\beta}}_{\lambda}^{(1)}=\hat{\bm{u}}_{n}^{(1)}+{\rm o}_{{\rm p}}(1)$ from Theorem \ref{thm1}, the terms including $\hat{\bm{\beta}}_{\lambda}^{(1)}$ do not reduce to ${\rm o}_{{\rm p}}(1)$ in this case. Therefore, (\ref{eq;taylor}) is expressed as
\begin{align*}
&\hat{\bm{u}}_{n}^{ (1){\rm T}}(\bm{s}_{n}^{(1)}-\bm{J}^{(12)}\bm{J}^{(22)-1}\bm{s}_{n}^{(2)})+(\bm{s}_{n}^{(2)}-\bm{p}'^{(2)}_{\lambda})^{{\rm T}}\bm{J}^{(22)-1}\bm{s}_{n}^{(2)} \\
&-\hat{\bm{u}}_{n}^{ (1){\rm T}}\bm{J}^{(1|2)}\hat{\bm{u}}_{n}/2-(\bm{s}_{n}^{(2)}-\bm{p}'^{(2)}_{\lambda})^{\rm T}\bm{J}^{(22)}(\bm{s}_{n}^{(2)}-\bm{p}'^{(2)}_{\lambda})/2+{\rm o}_{{\rm p}}(1),
\end{align*}
and this converges in distribution to
\begin{align*}
&\hat{\bm{u}}^{ (1){\rm T}}\bm{s}^{(1|2)}+(\bm{s}^{(2)}-\bm{p}'^{(2)}_{\lambda})^{{\rm T}}\bm{J}^{(22)-1}\bm{s}^{(2)} \\
&-\hat{\bm{u}}^{ (1){\rm T}}\bm{J}^{(1|2)}\hat{\bm{u}}/2-(\bm{s}^{(2)}-\bm{p}'^{(2)}_{\lambda})^{\rm T}\bm{J}^{(22)}(\bm{s}^{(2)}-\bm{p}'^{(2)}_{\lambda})/2.
\end{align*}
In the same way, (\ref{eq;taylor_copy}) is expressed as
\begin{align*}
&\hat{\bm{u}}_{n}^{ (1){\rm T}}(\tilde{\bm{s}}_{n}^{(1)}-\bm{J}^{(12)}\bm{J}^{(22)-1}\tilde{\bm{s}}_{n}^{(2)})+(\bm{s}_{n}^{(2)}-\bm{p}'^{(2)}_{\lambda})^{{\rm T}}\bm{J}^{(22)-1}\tilde{\bm{s}}_{n}^{(2)} \\
&-\hat{\bm{u}}_{n}^{ (1){\rm T}}\bm{J}^{(1|2)}\hat{\bm{u}}_{n}/2-(\bm{s}_{n}^{(2)}-\bm{p}'^{(2)}_{\lambda})^{\rm T}\bm{J}^{(22)}(\bm{s}_{n}^{(2)}-\bm{p}'^{(2)}_{\lambda})/2+{\rm o}_{{\rm p}}(1),
\end{align*}
and this converges in distribution to
\begin{align*}
&\hat{\bm{u}}^{ (1){\rm T}}\tilde{\bm{s}}^{(1|2)}+(\bm{s}^{(2)}-\bm{p}'^{(2)}_{\lambda})^{{\rm T}}\bm{J}^{(22)-1}\tilde{\bm{s}}^{(2)} \\
&-\hat{\bm{u}}^{ (1){\rm T}}\bm{J}^{(1|2)}\hat{\bm{u}}/2-(\bm{s}^{(2)}-\bm{p}'^{(2)}_{\lambda})^{\rm T}\bm{J}^{(22)}(\bm{s}^{(2)}-\bm{p}'^{(2)}_{\lambda})/2,
\end{align*}
where $\tilde{\bm{s}}_{n}^{(1)},\;\tilde{\bm{s}}^{(2)},\;\tilde{\bm{s}}^{(1|2)}$ and $\tilde{\bm{s}}^{(2)}$ are copies of $\bm{s}_{n}^{(1)},\;\bm{s}^{(2)},\;\bm{s}^{(1|2)}$ and $\bm{s}^{(2)}$, respectively. Thus, we see that
\begin{align*}
z^{{\rm limit}}
=\hat{\bm{u}}^{ (1){\rm T}}\bm{s}^{(1|2)}+(\bm{s}^{(2)}-\bm{p}'^{(2)}_{\lambda})^{{\rm T}}\bm{J}^{(22)-1}\bm{s}^{(2)}-\hat{\bm{u}}^{ (1){\rm T}}\tilde{\bm{s}}^{(1|2)}-(\bm{s}^{(2)}-\bm{p}'^{(2)}_{\lambda})^{{\rm T}}\bm{J}^{(22)-1}\tilde{\bm{s}}^{(2)}.
\end{align*}
Since $\tilde{\bm{s}}$ and $\bm{s}$ are independently distributed according to ${\rm N}(\bm{0},\bm{J}^{(22)})$, the asymptotic bias reduces to
\begin{align*}
{\rm E}[z^{{\rm limit}}]
={\rm E}[\hat{\bm{u}}^{ (1){\rm T}}\bm{s}^{(1|2)}]+{\rm E}[(\bm{s}^{(2)}-\bm{p}'^{(2)}_{\lambda})^{{\rm T}}\bm{J}^{(22)-1}\bm{s}^{(2)}].
\end{align*}
As a result, we obtain (\ref{eq;a1-3}).
\end{appendix}

\bibliography{ref_aic}

\end{document}